\documentclass{rQUF2e-arxiv}

\usepackage{epstopdf}
\usepackage{subfigure}

\theoremstyle{plain}

\theoremstyle{definition}

\theoremstyle{remark}

\usepackage[table, dvipsnames]{xcolor}
\usepackage{wasysym}
\usepackage{booktabs}
\usepackage{multirow}
\usepackage[group-separator={,}]{siunitx}
\usepackage{hyperref}

\newcommand{\greencell}[1]{{ \cellcolor{ForestGreen!10}} #1}
\newcommand{\avquant}[2]{\left \langle #1_{#2} \right \rangle}

\makeatletter
\newcommand{\tabhead}[1]{%
\multicolumn{1}{c}{#1}\checknextarg}
\newcommand{\checknextarg}{\@ifnextchar\bgroup{\gobblenextarg}{\\}}
\newcommand{\gobblenextarg}[1]{ & \multicolumn{1}{c}{#1}\@ifnextchar\bgroup{\gobblenextarg}{\\}}

\begin{document}


\title{Scaling of inefficiencies in the U.S.\ equity markets: Evidence from three market indices and more than 2900 securities}

\author{
John H. Ring IV$^{1,2,3,\star}$\thanks{$^\star$Corresponding author. Email: jhring@uvm.edu}
Colin M. Van Oort$^{4,3,2}$,
David R. Dewhurst$^{5,4,3}$,
Tyler J. Gray$^{3,4}$,
Christopher M. Danforth$^{4,3}$,
and Brian F. Tivnan$^{2,4,3,\star}$\thanks{$^\star$Corresponding author. Email: btivnan@mitre.org}
\affil{
$(1)$ Center for Computer Security and Privacy, University of Vermont, Burlington, VT 05405, USA\\
$(2)$ MITRE, McLean, VA 22012, USA\\
$(3)$ MITRE-UVM Computational Finance Lab, Burlington, VT 05405, USA\\
$(4)$ Vermont Complex Systems Center, University of Vermont, Burlington, VT 05405, USA\\
$(5)$ Charles River Analytics, Cambridge, MA 02368, USA
}
\received{October 2020}}

\maketitle

\begin{abstract}
Using the most comprehensive, commercially-available dataset of trading activity in U.S.\ equity markets, we catalog and analyze quote dislocations between the SIP National Best Bid and Offer (NBBO) and a synthetic BBO constructed from direct feeds.
We observe a total of over 3.1 billion dislocation segments in the Russell 3000 during trading in 2016, roughly 525 per second of trading.
However, these dislocations do not occur uniformly throughout the trading day.
We identify a characteristic structure that features more dislocations near the open and close.
Additionally, around 23\% of observed trades executed during dislocations.
These trades may have been impacted by stale information, leading to estimated opportunity costs on the order of \$2 billion USD\@.
A subset of the constituents of the S\&P 500 index experience the greatest amount of opportunity cost and appear to drive inefficiencies in other stocks.
These results quantify impacts of the physical structure of the U.S.\ National Market System.
\end{abstract}


\begin{classcode}D50, D82, G14\end{classcode}

\section{Introduction}\label{sec:intro}
Securities markets utilize auction mechanisms to facilitate the valuation and trade of assets~\citep{de1688confusion, bachelier1900theorie, knight1921risk, fama1965behavior}.
Implementation details of these markets, including the auction mechanism, computing and communication infrastructure, as well as information dissemination policies, impact their informational and economic efficiency~\citep{akerlof1978market, easley1987price}.
These market microstructure factors have been increasingly considered in recent analyses of market efficiency~\citep{wissner2010relativistic, ding2014slow, mackintosh2014need, adrian2016informational}, and we contribute to this body of work.

\subsection{Modern U.S.\ market}\label{subsec:modern-market}
We investigate a broad subset of the equities traded in the U.S.\ National Market System (NMS), a network of stock exchanges located in the U.S., since it is the proverbial center of the world equity markets.
In particular, we focus on constituents of the Russell 3000 Index, which is compiled by FTSE International Ltd.\ and contains roughly 3000 of the largest equities traded on the NMS\@.
The selected sample represents the vast majority of the equities traded in the U.S.\ and can serve as a nearly comprehensive cross-section of publicly traded equities from which the observation and assessment of microstructure quantities can be made.
\medskip

\begin{table}
    \tbl{
    Summary statistics of the realized opportunity cost (ROC) aggregated across all studied securities and all of calendar year 2016.
    The total ROC of this sample is over \$2B USD.
    We discuss statistical characteristics of ROC extensively in Section~\ref{sec:Results}.
    Row 10 shows that the average differing trade moves approximately 6.51\% more value than the average trade.
    This indicates a qualitative shift in trading behavior during dislocations.
    }
    {\begin{tabular}{llr}
        \toprule
        1  & Realized Opportunity Cost   &      \$2,051,916,739.66   \\
        2  & SIP Opportunity Cost        &      \$1,914,018,654.41   \\
        3  & Direct Opportunity Cost     &        \$137,898,085.25   \\
        4  & Trades                      &        4,745,033,119      \\
        5  & Diff.\ Trades               &        1,124,814,017      \\
        6  & Traded Value                & \$28,031,002,997,692.75   \\
        7  & Diff.\ Traded Value         &  \$7,077,357,462,641.67   \\
        8  & Percent Diff.\ Trades       &                   23.71   \\
        9  & Percent Diff.\ Traded Value &                   25.25   \\
        10 & Ratio of 9 / 8              &                    1.0651 \\
        \bottomrule
    \end{tabular}}
    \label{tab:all-nms-purse-table}
\end{table}

\begin{figure}
    \centering
    \includegraphics[width=0.8\textwidth]{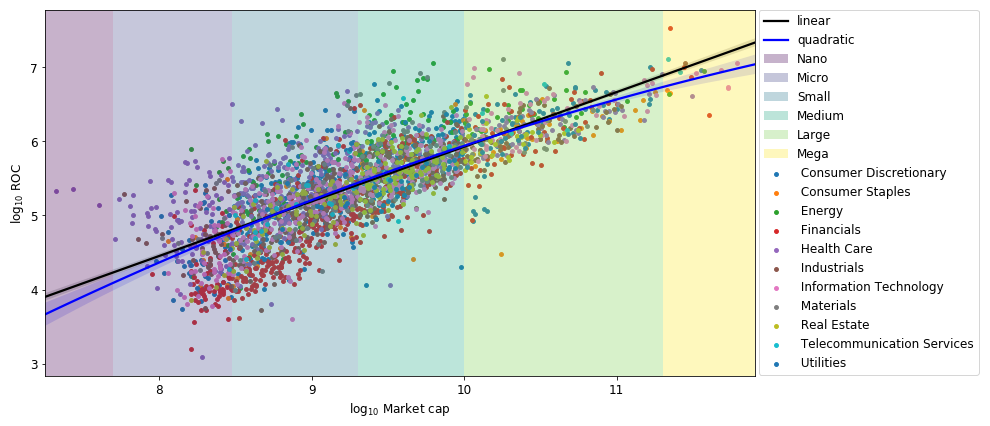}
    \caption{
    Linear and quadratic regression between Market Capitalization (MC) and ROC in doubly-logarithmic space.
    There is a strong positive relationship between MC and ROC.
    The data exhibits interesting nonlinearity and heteroskedasticity, where equities with smaller MC have higher variance in the dependent variable, while equities with larger MC have generally lower variance.
    Note that equities in the financial sector have a consistently lower ROC relative to MC while equities in the energy sector have a consistently higher ROC relative to MC.
    The shaded area surrounding the regression curves indicate 95\% confidence intervals for the true curves, calculated using bootstrapping techniques.
    }
    \label{fig:roc-mc-regression}
\end{figure}

\noindent
Using the most comprehensive dataset of NMS messages commercially available, we enumerate and describe dislocation segments (DS) and realized opportunity costs (ROC), defined in Section~\ref{sec:methods}, in Russell 3000 securities and a selection of exchange-traded funds (ETFs) during calendar year 2016.
Extending~\cite{tivnan2020fragmentation}, we observe over $3 \times 10^9$ DSs in the National Market System during the period of study.
Consistent with~\cite{tivnan2020fragmentation}, we define a dislocation as the market state when the National Best Bid and Offer (NBBO) and a synthetic, Direct Best Bid and Offer (DBBO) simultaneously display different prices.
Any trade that executes during a dislocation is called a differing trade.
Row 8 of Table~\ref{tab:all-nms-purse-table} shows that 23.71\% of all trades were differing trades.
Differing trades may have been influenced by stale quote information, so we used them to calculate ROC\@.
However, some trades may have been executed in this period intentionally, so we only include differing trades that executed at either of the two NBBO quotes.
This results in a conservative estimate of total ROC, \$2,051,916,739.66 across the Russell 3000 in 2016, as depicted in Row 1 of Table~\ref{tab:all-nms-purse-table}.
Table~\ref{tab:ds-number} depicts the total number of dislocations that we observed in 2016, delineating those that persisted for a minimum duration and those greater than the minimum magnitude (i.e., tick size).
Despite recent technological upgrades to market infrastructure, the chief economist at Nasdaq confirms that latencies of 500 microseconds remain material in 2020 for the execution of latency arbitrage strategies~\cite{mackintosh2020relativity}.

\begin{table}
    \tbl{
    Total number of dislocation segments in mutually-exclusive market categories.
    Number of opportunities is calculated unconditioned, conditioned on duration, and conditioned on both duration and magnitude.
    }
    {\begin{tabular}{lllr}
         \toprule
         \tabhead{Category}{Duration}{Magnitude}{Count}
         \midrule
         \multirow{3}{*}{Dow}       & -               & -                &   120,355,462 \\
         & $> 545 \mu s$     & -                &    65,073,196 \\
         & $> 545 \mu s$     & $>$ 1\cent       &     2,872,734 \\[.25em]
         \multirow{3}{*}{SPexDow}   & -               & -                & 1,126,186,592 \\
         & $> 545 \mu s$     & -                &   530,499,458 \\
         & $> 545 \mu s$     & $>$ 1\cent       &    51,187,430 \\[.25em]
         \multirow{3}{*}{RexSP}     & -               & -                & 1,888,686,248 \\
         & $> 545 \mu s$     & -                &   704,416,718 \\
         & $> 545 \mu s$     & $>$ 1\cent       &   110,447,787 \\
         \bottomrule
    \end{tabular}}
    \label{tab:ds-number}
\end{table}

\subsection{Scaling in finance}\label{subsec:scaling}
~\cite{mandelbrot1997variation, mandelbrot2013fractals} was one of the first to characterize the scaling properties of price returns in modern markets.
The scaling of returns was later revisited by ~\cite{stanley2001scaling}, ~\cite{cont2001empirical}, as well as ~\cite{patzelt2018universal}.
Beyond returns in price time series, additional financial variables have been found to display scaling properties.
Market indices and foreign exchange rates~\citep{di2007multi} as well as share volume and number of trades~\citep{stanley2008statistical} adhere to scaling properties.
\medskip

\begin{figure}[h]
    \centering
    \includegraphics[width=0.8\textwidth]{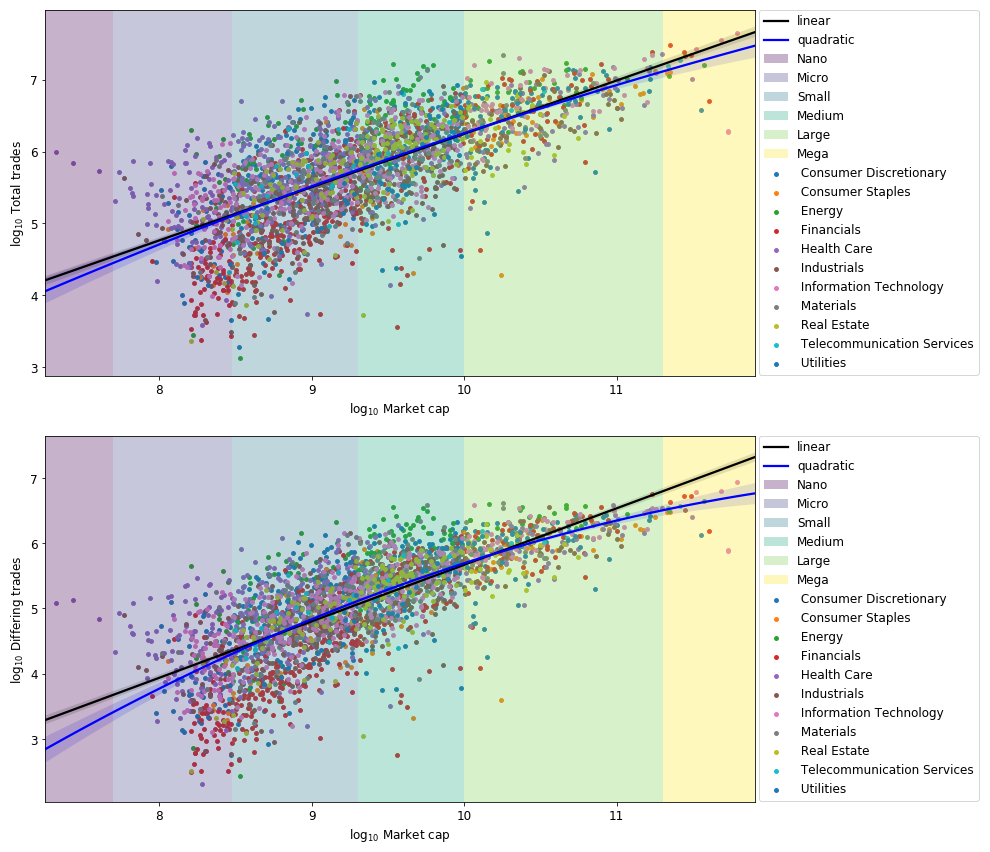}
    \caption{
    Relationships between Market Capitalization (MC) and total trades (top) or differing trades (bottom).
    Similar to Figure~\ref{fig:roc-mc-regression}, there is a strong positive relationship in both regressions, along with the same nonlinearity and heteroskedasticity.
    The data are well-fit by linear and quadratic functions in doubly-logarithmic space.
    The shaded area surrounding the regression curves indicate 95\% confidence intervals for the true curves, calculated using bootstrapping techniques.
    }
    \label{fig:alt-market-cap-regressions}
\end{figure}

\noindent
Leveraging the large number of securities under study and the broad range of market capitalization (MC) covered, we examine scaling relationships between DSs, ROC, and MC\@.
DSs occur in equities of all sizes.
While DS are more frequent in equities with larger MC, the distributions of their qualities, such as their magnitude and duration, are more extreme among equities with smaller MC\@.
We find a strong positive correlation between MC and ROC, show in Figure~\ref{fig:roc-mc-regression}.
A similar relationship is seen between MC-total trades and MC-differing trades in Figure~\ref{fig:alt-market-cap-regressions}.
The majority of ROC is generated by equities in the S\&P 500 that are not also in the Dow (termed the SPexDow).
The SPexDow also Granger-causes ROC in other mutually-exclusive market categories (Dow 30 and Russell 3000 less the S\&P 500, or RexSP), pointing to its centrality in the U.S.\ equities market.
\medskip

\noindent
In the following sections, we first provide a brief overview of the U.S.\ National Market System.
We then detail our data, the available and used fields, and summarize the equities studied.
After describing statistics of DSs, including distributions of start times and durations, we move to analysis of ROC, providing summary statistics, comparisons across mutually-exclusive market categories, and correlation along with Granger-causality analyses.
We close with a brief exploration of exchange-traded funds (ETFs), a discussion of results, and possibilities for future work.

\subsection{Empirical Studies of Modern U.S. Markets}\label{subsec:empirical-studies}
In a recent report to its government oversight committee, the U.S.\ Securities and Exchange Commission (SEC) offered the following characterization of the prevailing literature which relates to our study:
``It is unsurprising that academic studies generally are narrowly focused, as the amount of data, computing power and sophistication necessary to engage in broader study are daunting and costly, and relevant data may not be widely available or easily accessible.''~\hbox{\citep[p.\ 45]{sec2020algos}}.
Given these constraints, we are aware of only two other recent studies which also used comprehensive, market data to analyze the modern U.S.\ market.
\medskip

\noindent
In the first study, ~\cite{wah2016prevalent} calculated the potential opportunities for latency arbitrage on the S\&P 500 in 2014 using data from the SEC’s MIDAS platform~\citep{sec2013midas}.
Using similar data to that for our study, Wah identified price discrepancies that could serve as latency arbitrage opportunities.
She then examined the potential profit to be made by an infinitely fast arbitrageur taking advantage of these opportunities.
Wah estimates that this idealized arbitrageur could have captured \$3.0B USD from latency arbitrage in 2014, which is similar to our conservative calculations of approximately \$2.1B USD in ROC from actual trades in 2016.
\medskip

\noindent
The second study was ~\cite{aquilina2020quantifying}, which used message data from 2015 to quantify aspects of latency arbitrage in global equity markets.
The authors note the frequent yet fleeting occurrence of latency arbitrage opportunities and estimate profits from latency arbitrage in 2018 at \$4.8B USD globally, including \$2.8B USD in the U.S.\ equity market.
\medskip

\noindent
Both the Wah and Aquilina et al.\ studies relied on affiliations with regulatory agencies and their respective data.
This reliance on regulatory data supports the SEC observation that ``relevant data may not be widely available or easily accessible.''
\medskip

\section{Market Overview}\label{sec:market-overview}
The U.S.\ equities market, known as the National Market System (NMS), is composed of 13 National Securities Exchanges.
Each exchange contributes to price discovery through the interactions of market participants, mediated by an auction mechanism.
Another core component of the NMS is a collection of approximately 40 alternative trading systems (ATSs)~\citep{tuttle2013ats}, also known as dark pools.
ATSs provide limited pre-trade transparency, which can allow market participants to reduce the market impact of their trades, but have limited participation in price discovery as a result.
Each exchange and ATS accumulates orders whose execution conditions have not been met in an order book.
Resting orders are matched with incoming marketable orders based on a priority mechanism, commonly price-time priority~\citep{bessembinder2001price}.
Traders often have access to a variety of order types that allow them to tailor how they interact with the market~\citep{nasdaq_order_types,nyse_order_types,cboe_order_types,iex_order_types}.
The top of the book at each exchange, the resting bid with the highest price and the resting offer with the lowest price, is called the best bid and offer (BBO).
BBOs from across the NMS are aggregated by one of the Securities Information Processors (SIP) to form the national best bid and offer (NBBO)~\citep{cta_plan,utp_plan}.
Under Regulation National Market System (Reg.\ NMS), trades must execute at a price that is no worse than the NBBO, though exceptions exist (e.g.~intermarket sweep orders)~\citep{reg_nms}.
\medskip

\noindent
Market participants in the NMS have several options of data products to fuel their trading decisions.
In addition to the dissemination of the NBBO, each SIP provides data feeds containing all quotes and trades that occur in their managed securities.
Information feeds offered by each exchange, referred to as direct feeds, can provide similar information with lower latency than the SIP data feeds.
Direct feeds can also provide additional data, such as the resting volume at all price points, commonly called depth-of-book information.
Information asymmetries between data products lead to DSs, which can impact trading decisions and outcomes.
\medskip

\noindent
The NMS is regulated by the U.S.\ Securities and Exchange Commission (SEC), a federal agency, and self-regulated by the Financial Industry Regulatory Authority (FINRA), a professional organization.
FINRA polices its members and ensures they adhere to SEC rules and other professional guidelines, while SEC designs, implements, and enforces rules that are intended to promote market stability and economic efficiency.
The physical structure of the NMS, in conjunction with the existence and usage of multiple distinct information feeds, leads to the creation of DSs and associated ROC\@.
On Dow 30 equities, over 120 million DSs and over \$160M USD in ROC were cataloged during calendar year 2016~\citep{tivnan2020fragmentation}.
\medskip

\noindent
Our calculations provide a conservative estimate of ROC from actual trades in the U.S.\ equity markets in 2016.
Therefore, we identify some relevant literature on trade execution~\citep{sec2013execution}; namely, where and when trades occur.
First, trading is not instantaneous.
Delays, or latencies, exist throughout the NMS\@.
Second, not all trading activity occurs at a national exchange or an ATS\@.
Instead of routing an order to one of these market venues, a broker may execute the order against the broker’s own inventory of that stock.
This process of retaining customers’ orders internal to the brokerage is called ``internalization"~\citep{sec2000internalization}.
In addition to matching customers’ orders against the broker’s inventory of a particular stock, internalization also includes instances when a broker may route customers’ orders to a market-maker under a Payment for Order Flow (PFOF) agreement.
Even without charging commissions for trades, brokers may generate revenue from executing trades via PFOF~\citep{sec2007pfof}.
To mitigate potential conflicts of interest, each broker is required to ensure that its customers’ orders execute against best prices, as determined by the NBBO\@.
\medskip

\noindent
Trade execution problems may still arise from PFOF\@.
In a public statement announcing its fine against a prominent market-maker, the SEC noted the use of algorithms which were used to avoid paying best prices on internalized orders.
Per the SEC, ``these algorithms were triggered when they identified differences in the best prices on market feeds, comparing the SIP feeds to the direct feeds from exchanges”~\citep{sec2017citadel}.
The reader will note that this market state, what the SEC has identified as “differences in best prices on market feeds,” is the very same state that we have defined here as a market dislocation.
\medskip

\noindent
PFOF remains a controversial practice.
More recently, another market-maker settled allegations that it did not ensure best prices for the internalization of its customers’ orders~\citep{michaels2019robinhood}.
\medskip

\noindent
We found references to internalization and PFOF dating back to 1994, when annual revenues from PFOF exceeded \$500M USD across all U.S.\ brokers~\citep{wsj1994inhouse}.
Some studies identified the potential for conflicts of interest from PFOF, but claimed that these conflicts could be mitigated by the adoption of minimum tick sizes of a penny (i.e., decimalization)~\citep{chordia1995market, easley1996cream}.
Though the SEC adopted decimalization in 2000~\citep{sec2000decimal}, PFOF remains a lucrative practice.
In the first half of 2020, four brokers in the U.S.\ generated more than \$1B USD in revenue from PFOF~\citep{wursthorn2020robinhood}.
\medskip

\noindent
This brief overview of the U.S.\ equities market only provides context for the following sections and is far from complete.
We refer the reader to~\cite{tivnan2020fragmentation} for a more complete discussion.

\section{Indices}\label{sec:indices}
Many of our results are centered around the components of three of the most popular equity indices: Dow Jones Industrial Average, S\&P 500, and the Russell 3000.
Indices measure the performance of a bucket of securities.
The choice of the underlying securities is often to be representative of a market segment.
Indices may not be directly purchased in the same way as an equity, but may be tracked by Exchange Traded Funds (ETFs) and mutual funds.

The Dow Jones Industrial Average, from here on referred to as the Dow, is a price weighted index that aims to provide an overview of the U.S.\ economy~\citep{dowmethods}.
The Dow consists of thirty S\&P 500 constituents, covering all industries except for utilities and transportation.

The S\&P 500 is a market capitalization weighted index of 500 large US based companies referred to by it's creators as ``the gauge of the market economy".
The index is considered by many to be representative of the US stock market as a whole and is a primary holding among passive investors.
To be included in the index, as of 2016, a company must meet the following criteria~\citep{sp500methods}.
\begin{itemize}
    \item Be a U.S.\ Company
    \item Have a market capitalization greater than \$5.3 billion
    \item Be highly liquid
    \item Have a public float of at least 50\% of outstanding shares
    \item Had positive earnings in the most recent quarter
    \item The sum of the last four consecutive quarterly earnings must be positive
    \item Be listed on a major exchange
\end{itemize}
Meeting these criteria does not guarantee inclusion, and failing to uphold these standards does not necessarily result in immediate expulsion from the index.
S\&P 500 constituents are chosen by S\&P Global, and the index is updated regularly, though not on any fixed schedule.

The Russell indexes are passively constructed (no human in the loop) based on a transparent set of rules including~\citep{russmethods}:
\begin{itemize}
    \item Be a U.S.\ Company
    \item Be listed on a major exchange
    \item Have a share price $\ge$ \$1
    \item Have a market capitalization $\ge$ \$30M
    \item Have a public float $\ge$ 5\%
\end{itemize}
The Russell 3000 consists of the largest 3000 firms by market capitalization meeting the above criteria, or the entire eligible set, whichever is smaller.
The index undergoes an annual reconstruction in June and is augmented quarterly with the addition of Initial Public Offerings (IPO).
This methodology results in the Russell 3000 being a strict superset of the S\&P 500.

For our analysis we focus on constituents of these indices, rather the index itself.
Thus, differing weighting methodologies used by these indices have no effect on our analysis.
We also note that some companies have multiple common stocks, one for each share class, and that each index handles the inclusion of multiple share classes differently.

\section{Data}\label{sec:data}
We use a dataset comprised of every quote and trade message that was disseminated on one of the SIP or direct feeds during the period of study.
This dataset features comprehensive coverage of the stocks under study, is collected from a single location (Carteret, NJ), and is time stamped upon arrival, thus limiting clock synchronization issues.
Thesys Technologies collected and curated this data~\citep{thesys_tech}, and also provided data for the SEC's MIDAS~\citep{sec2013midas} at the time of collection.
Index membership, Global Industry Classification Standard (GICS) sector classification, and MC data were obtained from a Bloomberg Terminal.
\medskip

The indices we consider are subject to frequent changes in membership.
To simplify our analysis we consider the Dow 30 and S\&P 500 as they stood on Jan.\ 1, 2016.
For the Russell 3000 we consider the constituents as listed in the June 2016 construction, excluding those that were not publicly traded on Jan.\ 1, 2016.
Constituents of the indices under study were curated to only include companies that survived as a publicly traded entity on a national exchange for the entire calendar year of 2016.
Companies that were delisted for any reason (e.g.\ bankruptcy or buyout) were excluded, in addition to those who were acquired by an out-of-study firm.
Mergers between in-sample companies did not result in exclusion.
Curating the stocks under study in this way allows us to avoid issues caused by IPOs and delistings.
\medskip

\noindent
Many companies in our dataset changed their ticker symbol over the course of the calendar year and thus appear as a different entity in the data.
To study a company over a long time period it is necessary to know all tickers it traded under and when the ticker changes occurred.
There is no consolidated public record of these ticker changes, so we tracked them via an extensive review of press releases.
These ticker changes were then validated by observing changes in trading activity in the old and new ticker on the date of the change using the Thesys data archive.
\medskip

This curation reduced the Russell 3000 from 3005 stocks to 2903, the S\&P 500 from 500 stocks to 472, and did not impact the 30 members of the Dow.
We denote the curated version of an existing index by appending a prime to the respective base index (e.g.\ Dow 30 $\rightarrow$ Dow 30$^{\prime}$).
We then construct two additional stock groups, RexSP and SPexDow, by taking the appropriate set difference, e.g.\ SPexDow = S\&P 500$^{\prime}$ - Dow 30$^{\prime}$.
Finally, all companies in our dataset were classified by their MC as it stood in the beginning of Q4 2016 using the classes defined in Table~\ref{tab:market_cap_classes}.
Our dataset covers approximately 98\% of all publicly traded U.S.\ equities by MC~\citep{r3k_factsheet}.
Tables~\ref{tab:sectors} -~\ref{tab:market-cap} provide summary statistics and distribution of these equities across GICS sector, MC, and market category, for several indices.

\begin{table}
    \tbl{
    Composition of indexes under study by market capitalization (MC) classification as of Q4 2016.
    The composition of various indexes is displayed by the percentage of index constituents that are a member of each given index (\% by \#) and by the weighting of those constituents (\% by MC).
    }
    {\begin{tabular}{llrrrrr}
        \toprule
        \tabhead{Class}{Statistic}{Russ 3K$^\prime$}{RexSP}{S\&P 500$^\prime$}{SPexDow}{Dow 30$^\prime$}
        \midrule
        \multirow{3}{*}{Nano} & \% by \# &              0.14 &   0.16 &               0.00 &     0.00 &             0.00 \\
        & \% by MC &              0.00 &   0.00 &               0.00 &     0.00 &             0.00 \\
        & Count &                 4 &      4 &               0 &     0 &             0 \\
        \midrule
        \multirow{3}{*}{Micro} & \% by \# &             11.51 &  13.74 &               0.00 &     0.00 &             0.00 \\
        & \% by MC &              0.26 &   1.09 &               0.00 &     0.00 &             0.00 \\
        & Count &               334 &    334 &               0 &     0 &             0 \\
        \midrule
        \multirow{3}{*}{Small} & \% by \# &             42.89 &  51.13 &               0.42 &     0.45 &             0.00 \\
        & \% by MC &              4.37 &  18.50 &               0.01 &     0.02 &             0.00 \\
        & Count &             1,245 &  1,243 &                  2 &        2 &             0 \\
        \midrule
        \multirow{3}{*}{Mid} & \% by \# &             30.35 &  32.21 &              20.76 &    22.17 &             0.00 \\
        & \% by MC &             15.11 &  53.19 &               3.37 &     4.72 &             0.00 \\
        & Count &               881 &    783 &                 98 &       98 &             0 \\
        \midrule
        \multirow{3}{*}{Large} & \% by \# &             14.50 &   2.71 &              75.21 &    75.79 &            66.67 \\
        & \% by MC &             56.68 &  20.72 &              67.77 &    77.59 &            43.28 \\
        & Count &               421 &     66 &                355 &      335 &               20 \\
        \midrule
        \multirow{3}{*}{Mega} & \% by \# &              0.62 &   0.04 &               3.60 &     1.58 &            33.33 \\
        & \% by MC &             23.58 &   6.50 &              28.85 &    17.67 &            56.72 \\
        & Count &                18 &      1 &                 17 &        7 &               10 \\
        \bottomrule
    \end{tabular}}
    \label{tab:market_cap_classes}
\end{table}

\begin{table}
    \tbl{
    Market Capitalization (MC) statistics of equities under study broken out by Global Industry Classification Standard (GICS) sector as of Q4 2016.
    The composition of various indexes is displayed by the percentage of index constituents that are a member of each given sector (\% by \#) and by the weighting of those constituents (\% by MC).
    Additionally, the MC of the smallest and largest constituent for each index in each category is displayed.
    }
    {\begin{tabular}{clrrrrr}
        \toprule
        \tabhead{Sector}{Statistic}{Russ 3K$^\prime$}{RexSP}{S\&P 500$^\prime$}{SPexDow}{Dow 30$^\prime$}
        \midrule
        \multirow{5}{*}{\shortstack{Consumer\\ Discretionary}} & \% by \# &             14.92 &           14.52 &              16.95 &           17.19 &            13.33 \\
        & \% by MC &             12.97 &           16.40 &              11.92 &           13.10 &             8.98 \\
        & Count &               433 &             353 &                 80 &              76 &                4 \\
        &  (\$) MC Min &        95,330,024 &      95,330,024 &      1,244,719,232 &   1,244,719,232 &   84,654,022,656 \\
        &  (\$) MC Max &   356,313,137,152 &  89,539,158,016 &    356,313,137,152 & 356,313,137,152 &  165,862,064,128 \\
        \midrule
        \multirow{5}{*}{\shortstack{Consumer\\ Staples}} & \% by \# &              4.03 &            3.41 &               7.20 &            7.01 &               10 \\
        & \% by MC &              8.54 &            3.83 &               9.99 &            9.69 &            10.74 \\
        & Count &               117 &              83 &                 34 &              31 &                3 \\
        &  (\$) MC Min &       114,570,432 &     114,570,432 &      9,794,159,616 &   9,794,159,616 &  178,815,287,296 \\
        &  (\$) MC Max &   224,997,457,920 &  17,508,790,272 &    224,997,457,920 & 150,058,582,016 &  224,997,457,920 \\
        \midrule
        \multirow{5}{*}{Energy} & \% by \# &              5.20 &            4.73 &               7.63 &            7.69 &             6.67 \\
        & \% by MC &              6.57 &            4.71 &               7.14 &            5.83 &            10.40 \\
        & Count &               151 &             115 &                 36 &              34 &                2 \\
        &  (\$) MC Min &       160,502,160 &     160,502,160 &      2,427,903,232 &   2,427,903,232 &  222,190,436,352 \\
        &  (\$) MC Max &   374,280,552,448 &  27,468,929,024 &    374,280,552,448 & 116,800,331,776 &  374,280,552,448 \\
        \midrule
        \multirow{5}{*}{Financials} & \% by \# &             17.81 &           18.84 &              12.50 &           12.44 &            13.33 \\
        & \% by MC &             15.17 &           21.99 &              13.07 &           14.73 &             8.91 \\
        & Count &               517 &             458 &                 59 &              55 &                4 \\
        &  (\$) MC Min &        89,903,488 &      89,903,488 &      3,021,111,552 &   3,021,111,552 &   34,774,474,752 \\
        &  (\$) MC Max &   401,644,421,120 & 401,644,421,120 &    308,768,440,320 & 276,779,139,072 &  308,768,440,320 \\
        \midrule
        \multirow{5}{*}{Health Care} & \% by \# &             15.23 &           15.84 &              12.08 &           11.99 &            13.33 \\
        & \% by MC &             12.49 &            9.12 &              13.53 &           13.19 &            14.38 \\
        & Count &               442 &             385 &                 57 &              53 &                4 \\
        &  (\$) MC Min &        21,050,850 &      21,050,850 &      1,478,593,408 &   1,478,593,408 &  152,328,667,136 \\
        &  (\$) MC Max &   313,432,473,600 &  18,889,377,792 &    313,432,473,600 & 108,768,911,360 &  313,432,473,600 \\
        \midrule
        \multirow{5}{*}{Industrials} & \% by \# &             13.47 &           13.41 &              13.77 &           13.57 &            16.67 \\
        & \% by MC &             10.40 &           11.03 &              10.20 &            9.91 &            10.94 \\
        & Count &               391 &             326 &                 65 &              60 &                5 \\
        &  (\$) MC Min &        58,695,636 &      58,695,636 &      2,821,674,240 &   2,821,674,240 &   54,259,630,080 \\
        &  (\$) MC Max &   279,545,937,920 &  13,281,452,032 &    279,545,937,920 & 100,041,220,096 &  279,545,937,920 \\
        \midrule
        \multirow{5}{*}{\shortstack{Information\\ Technology}} & \% by \# &             14.40 &           14.60 &              13.35 &           12.90 &               20 \\
        & \% by MC &             21.40 &           13.81 &              23.74 &           20.93 &            30.74 \\
        & Count &               418 &             355 &                 63 &              57 &                6 \\
        &  (\$) MC Min &       114,370,240 &     114,370,240 &      3,334,570,240 &   3,334,570,240 &  151,697,113,088 \\
        &  (\$) MC Max &   617,588,457,472 &  32,402,583,552 &    617,588,457,472 & 538,572,161,024 &  617,588,457,472 \\
        \midrule
        \multirow{5}{*}{Materials} & \% by \# &              4.55 &            4.40 &               5.30 &            5.43 &             3.33 \\
        & \% by MC &              3.26 &            5.83 &               2.47 &            3.02 &             1.11 \\
        & Count &               132 &             107 &                 25 &              24 &                1 \\
        &  (\$) MC Min &       103,733,456 &     103,733,456 &      2,823,849,728 &   2,823,849,728 &   63,809,703,936 \\
        &  (\$) MC Max &    69,704,540,160 &  69,704,540,160 &     63,809,703,936 &  46,132,944,896 &   63,809,703,936 \\
        \midrule
        \multirow{5}{*}{Real Estate} & \% by \# &              6.61 &            6.99 &               4.66 &            4.98 &             0.00 \\
        & \% by MC &              3.89 &            8.67 &               2.41 &            3.38 &             0.00 \\
        & Count &               192 &             170 &                 22 &              22 &             0 \\
        &  (\$) MC Min &       161,591,616 &     161,591,616 &      7,130,559,488 &   7,130,559,488 &             0.00 \\
        &  (\$) MC Max &    55,830,577,152 &  24,264,243,200 &     55,830,577,152 &  55,830,577,152 &             0.00 \\
        \midrule
        \multirow{5}{*}{\shortstack{Telecommunication\\ Services}} & \% by \# &              1.03 &            1.03 &               1.06 &            0.90 &             3.33 \\
        & \% by MC &              2.40 &            1.82 &               2.57 &            2.09 &             3.79 \\
        & Count &                30 &              25 &                  5 &               4 &                1 \\
        &  (\$) MC Min &       285,299,072 &     285,299,072 &      3,964,831,488 &   3,964,831,488 &  217,610,731,520 \\
        &  (\$) MC Max &   261,176,721,408 &  47,389,126,656 &    261,176,721,408 & 261,176,721,408 &  217,610,731,520 \\
        \midrule
        \multirow{5}{*}{Utilities} & \% by \# &              2.76 &            2.22 &               5.51 &            5.88 &             0.00 \\
        & \% by MC &              2.91 &            2.78 &               2.95 &            4.13 &             0.00 \\
        & Count &                80 &              54 &                 26 &              26 &             0 \\
        &  (\$) MC Min &       141,720,064 &     141,720,064 &      3,867,331,328 &   3,867,331,328 &             0.00 \\
        &  (\$) MC Max &    57,253,351,424 &  12,880,323,584 &     57,253,351,424 &  57,253,351,424 &             0.00 \\
        \bottomrule
    \end{tabular}}
    \label{tab:sectors}
\end{table}

\begin{table}
    \tbl{
    Makeup of market indexes by number of constituents as of Q4 2016.
    Additionally, the Market Capitalization (MC) of the smallest and largest constituent for each index is displayed along with the sum of all constituent MCs.
    }
    {\begin{tabular}{lrrrrr}
        \toprule
        \tabhead{}{Russ 3K$^\prime$}{RexSP}{S\&P 500$^\prime$}{SPexDow}{Dow 30$^\prime$}
        \midrule
        Count       &              2,903 &             2,431 &                472 &                442 &                30\\
        (\$) MC Sum & 26,217,754,755,404 & 6,177,292,648,268 & 20,040,462,107,136 & 14,303,673,004,544 & 5,736,789,102,592 \\
        (\$) MC Min &         21,050,850 &        21,050,850 &      1,244,719,232 &      1,244,719,232 &    34,774,474,752 \\
        (\$) MC Max &    617,588,457,472 &   401,644,421,120 &    617,588,457,472 &    538,572,161,024 &   617,588,457,472 \\
        \bottomrule
    \end{tabular}}
    \label{tab:market-cap}
\end{table}

\begin{figure}
    \centering
    \includegraphics[width=.8\textwidth]{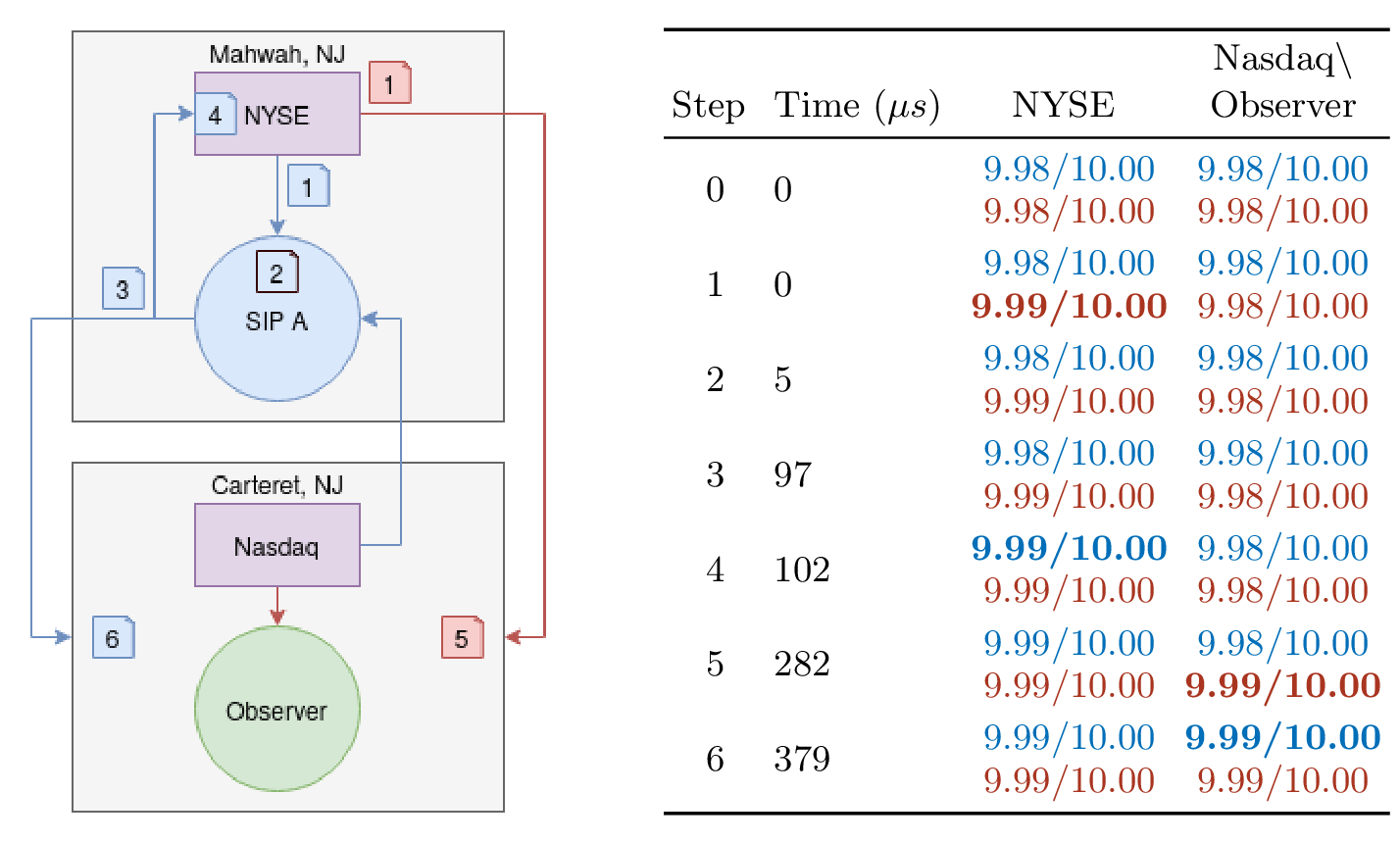}
    \caption{
    We depict the dissemination of a market event to a subset of core participants in the national market system.
    The left panel visualizes the the plumbing connecting our participants;
    NYSE and SIP tape A co-located in Mahwah, NJ and Nasdaq along with our observer co-located in Carteret, NJ.
    All participants subscribe to both the SIP (blue) and direct feeds (red) from both exchanges.
    We show the flow of information as a sequence of enumerated events depicted as rectangular documents.
    The right panel displays the best bid and offer observed by the participants at each event from both the SIP (blue) and direct feeds (red).
    Note that while Nasdaq and our observer remain in sync for this entire example this is not always the case.
    We start at step zero with a market in harmony, that is all participants observe the same price on all feeds.
    Within the same microsecond NYSE processes an order resulting in a new best bid that narrows the spread.
    NYSE quickly dispatches a message of the top-of-book change to the SIP and its direct feed customers.
    Five microseconds later~\citep{bats_colo, cboe_colo} NYSE's message arrives at the SIP which takes an additional $92\mu s$~\citep{cta_metrics} to process the information and dispatch a new NBBO.
    After another five microseconds NYSE receives the new NBBO from its co-located SIP.
    It's not for another $180\mu s$, $282\mu s$ after the original message the subscribers to NYSE's direct feed in Carteret receive the message.
    At this point we observe a 1\cent\ dislocation between the BBO displayed on the direct feeds and the observed NBBO.
    This dislocation persists for $97\mu s$ at which time the SIP update arrives in Carteret.
    Note that while technological advances will result in this sequence of events unfolding faster, the core behavior will remain unchanged.
    Messages from direct feeds travel a single leg, from exchange to subscriber, while updates to the NBBO require two legs, exchange to SIP to subscriber.
    }
    \label{fig:dislocation_example}
\end{figure}

\section{Methods}\label{sec:methods}
Our work investigates the occurrence of DSs and ROC arising from quote discrepancies between the SIP and direct feeds.
Similar concepts have been discussed in empirical market microstructure literature~\citep{arnuk2009latency, jarrow2012dysfunctional, hasbrouck2013low, ding2014slow, o2015high}, though formal definitions vary.
We follow the definitions described in ~\cite{tivnan2020fragmentation}, which are briefly reviewed below.
\medskip

\noindent
Suppose that there exist two market data feeds, $F_1$ and $F_2$, each displaying quotes for a single asset.
Quotes have the form $q_i(t) = (b_i(t), m_i(t), o_i(t), n_i(t))$, where $i \in {1, 2}$, $b_i(t)$ is the bid price at time $t$, $o_i(t)$ is the offer priceat time $t$, $m_i(t)$ and $n_i(t)$ are the number of shares associated with the bid and offer at time $t$ respectively.
We observe these feeds from a single, fixed location in Carteret, NJ\@.
A dislocation between these sources of data occurs when the prices of the quotes differ, e.g.\ $b_1(t) \neq b_2(t)$ or $o_1(t) \neq o_2(t)$.
A DS occurs when the quotes differ and the relationship between the quoted prices remains constant, e.g.\ $b_1(t) < b_2(t)$ or $b_1(t) > b_2(t)$.
Figure~\ref{fig:dislocation_example} walks through an example DS occurring on a subset of the NMS using estimates of message transit and processing time for each leg of the journey.
In our example a DS starts when a message regarding a quote change in Mahwah reaches our observer in Carteret via a direct feed and ends when the same message arrives via the SIP $92\mu s$ later.
In this single example we see three factors that either alone, or in combination, may cause DSs;
differences in processing time, transfer speed, and route (SIP messages require an additional leg).
In this example the dislocation was triggered by a single top of book change at NYSE\@.
However, dislocations can occur due to sequences of events occurring across multiple exchanges and SIP processors.
Recall that by definition a DS requires two feeds.
TAQ data contains only the quotes resulting in a NBBO change as well as all trades.
In contrast our dataset contains all quotes sent along the direct feeds as well as all SIP updates.
Thus, we can observe events such as our example in Figure~\ref{fig:dislocation_example}, an impossibility with TAQ data.
\medskip

\noindent
The ROC of using $F_1$ instead of $F_2$ is calculated by combining quote and trade information.
Assume that trades take the form $T_j = (p_j, v_j, t_j)$, where $p_j$ is the execution price, $v_j$ is the number of traded shares, and $t_j$ is the execution time.
If a trade executes at one of the currently quoted prices, e.g.\ $b_1(t_j)$, then the ROC is given by $(b_2(t_j) - b_1(t_j)) * v_j$.
If the trade executes on the opposite side of the book, e.g.\ $o_2(t_j)$, then the ROC is given by $(o_1(t_j) - o_2(t_j)) * v_j$.
This allows for a consistent interpretation of the values, where a positive value indicates that $F_2$ displayed a better price for the active trader (higher bid or lower offer) than $F_1$.
The total ROC over an interval $[S,\ E]$ is obtained by taking the sum of ROC values from all trades that occurred in that interval.
\medskip

\noindent
We first compute summary statistics and qualitative descriptions of the distributions of DSs and ROC\@.
Additionally, we leverage the large sample of equities to conduct a cross-sectional study of the effect of company ``size'' on these microstructure quantities.
We quantify the notion of size of a company by both its MC and its rank in relation to other companies.
We also investigate index inclusion effects through the use of disjoint sets of equities and compute aggregate statistics across these sets.
Since the S\&P 500 is a strict superset of the Dow 30 and the Russell 3000 is a strict superset of the S\&P 500, the natural division of the superset of all equities under study is split into three distinct classes:
the Dow 30, the S\&P 500 excluding the Dow 30 (SPexDOW), and the Russell 3000 excluding the S\&P 500 (RexSP).
We investigate correlations between these disjoint subsets, and characterize the statistical properties of the time series of DSs and ROC across these disjoint categories.
We further explore the relationship between these categories by conducting a Granger causality analysis of aggregated ROC time series~\citep{granger1969investigating}.
\medskip

\noindent
The next section gives results on DSs, including summary statistics and regressions of DSs against MC\@.
We then discuss structure in the intra-day distribution of DS start times and DS duration.
Following this, we provide statistics of the ROC across the market as a whole and again within mutually-exclusive market categories.
We then explore statistical properties of the ROC time series.
We close with an overview of the statistics of ETF DSs and ROC, contrasting these with those of the market as a whole.

\section{Results}\label{sec:Results}
\subsection{Dislocation Segments}\label{subsec:dislocation-segments}

DSs can occur when quotes displayed by distinct information feeds differ.
We cataloged all DSs occurring in the equities under study and present summary statistics along with qualitative comparisons of their distributions and higher-order moment statistics.
Table~\ref{tab:ds-dow} -~\ref{tab:ds-rexsp} display means of summary statistics of DSs for each mutually-exclusive market category under study.

\begin{figure}
    \centering
    \includegraphics[width=.8\textwidth]{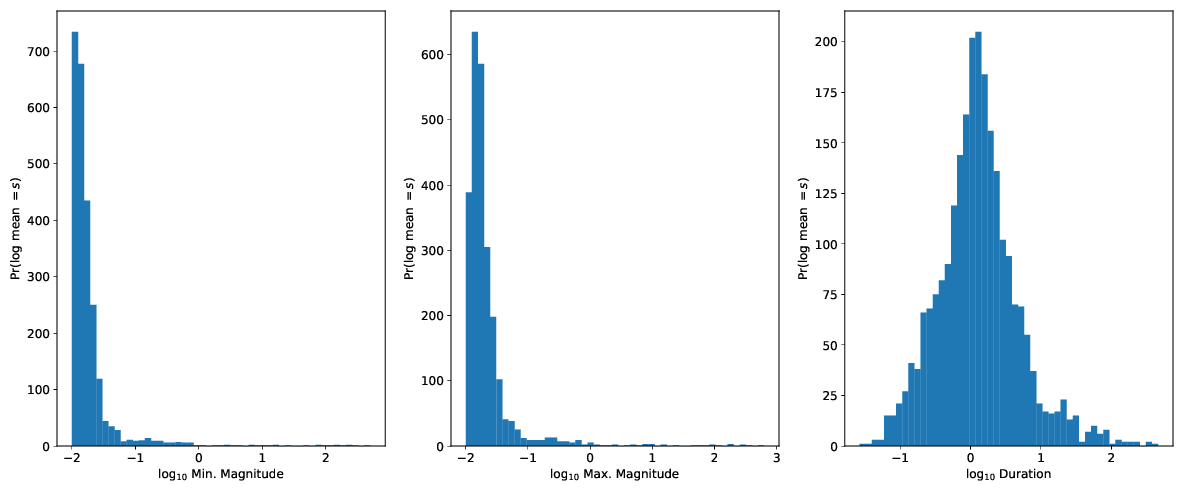}
    \caption{
    Histograms of the base-10 logarithm of minimum magnitude, maximum magnitude, and duration of dislocation segments in the RexSP without conditioning on duration or magnitude.
    The distributions are leptokurtic, with the log-distributions of minimum and maximum magnitude presenting a long right tail and the distribution of log-duration displaying a rough bell-shape.
    }
    \label{fig:ds-rexsp-example}
\end{figure}

We will use the notation $\avquant{f}{A}$ to denote an average of the quantity $f$ conditioned on the condition $A$.
These averages are interpreted as the quantity $f$ conditioned on condition $A$ averaged over all securities and all times of observation;
defining the number of instances of the quantity $f$ having condition $A$ as $N_A$, we have
\begin{equation}
    \avquant{f}{A} = \frac{1}{N_A}\sum_{
    \substack{1 \leq n \leq N_A \\ f \text{ has condition }A}
    }f_n.
    \label{eq:1}
\end{equation}
Tables~\ref{tab:ds-dow} -~\ref{tab:ds-rexsp} show that, on average, there were more DSs in Dow 30 securities than in SPexDow or RexSP securities.
However, the average maximum magnitude of DSs in the Dow30 is lower than those of the SPexDow, which in turn are lower than those of the RexSP\@.
In particular, actionable DSs (those with duration $> 545 \mu s$) with magnitude $>$ \$0.01 exhibit more extreme behavior in the SPexDow and RexSP than in the Dow.
On average, the median maximum magnitude in the Dow 30 among actionable DSs was $\avquant{\text{median max mag}}{duration, magnitude} \simeq \$0.023$, while in the SPexDow we observed $\avquant{\text{median max mag}}{duration, magnitude} \simeq \$0.034$ and in the RexSP $\avquant{\text{median max mag}}{duration, magnitude} \simeq \$0.045$, a roughly one-cent increase in the median maximum magnitude of a DS in each mutually-exclusive market category.
Examples of distributions of these quantities are given in Figure~\ref{fig:ds-rexsp-example}, where the distributions of the means of minimum magnitude, maximum magnitude, and duration are plotted for the RexSP\@.
\medskip

\noindent
These results provide evidence for the existence of a MC scaling effect in DSs.
Securities with larger MC tend to feature higher trading volume and more frequent occurrence of DSs, but these DSs tend to be smaller in magnitude on average.
More frequent trading implies a lower probability that prices across differing information feeds will diverge by large magnitudes.
\medskip

\noindent
Since DSs are not distributed evenly throughout the day in the Dow 30~\citep{tivnan2020fragmentation}, we examine their distribution in the SPexDow and the RexSP as well.
Appendix~\ref{sec:figures} contains figures displaying the distribution of DS start times plotted modulo day and aggregated over the year as well as figures displaying the distribution of DS durations for each mutually exclusive market category.
Distributions are plotted both without conditioning, conditioned on duration, as well as conditioned on duration and magnitude.
\medskip

\noindent
Distributions of start times display predictable structure.
In all market categories, there are large peaks at the very beginning and end of the trading day (circa 9:30 AM and 4:00 PM), along with a noticeable and sudden increase in density around 2:00 PM\@.
The peak in density that occurs at the end of the day is most noticeable when the distribution of start times is not conditioned on DS size.
These observations correspond with the results found for the Dow 30 in~\cite{tivnan2020fragmentation}.
However, along with these granular observations, there also exists structure on shorter timescales.
The distribution exhibits self-similarity on the half-hour timescale, with large peaks every half-hour and decreasing density toward a sudden peak at the next half-hour.
There is also structure at the five-minute timescale that is noticeable before the 2:00 PM spike in density but does not appear to be present after the spike.
(Future work could statistically test for the presence of this structure and for its persistence across multiple timescales.)
The structure on shorter timescales is present in all distributions but, again, is more pronounced in distributions not conditioned on magnitude.
\medskip

\noindent
Distributions of DS duration are extremely heavy tailed, so we plot them with a log-transformed horizontal axis.
All DS duration distributions exhibit one or more peaks in the range $10^{-4}s \leq \log_{10} \text{duration} \leq 10^{-3}s$, but there is also a distinct and much lower peak in the distribution near approximately one second in length.
\medskip

\begin{figure}
    \centering
    \includegraphics[width=0.8\textwidth]{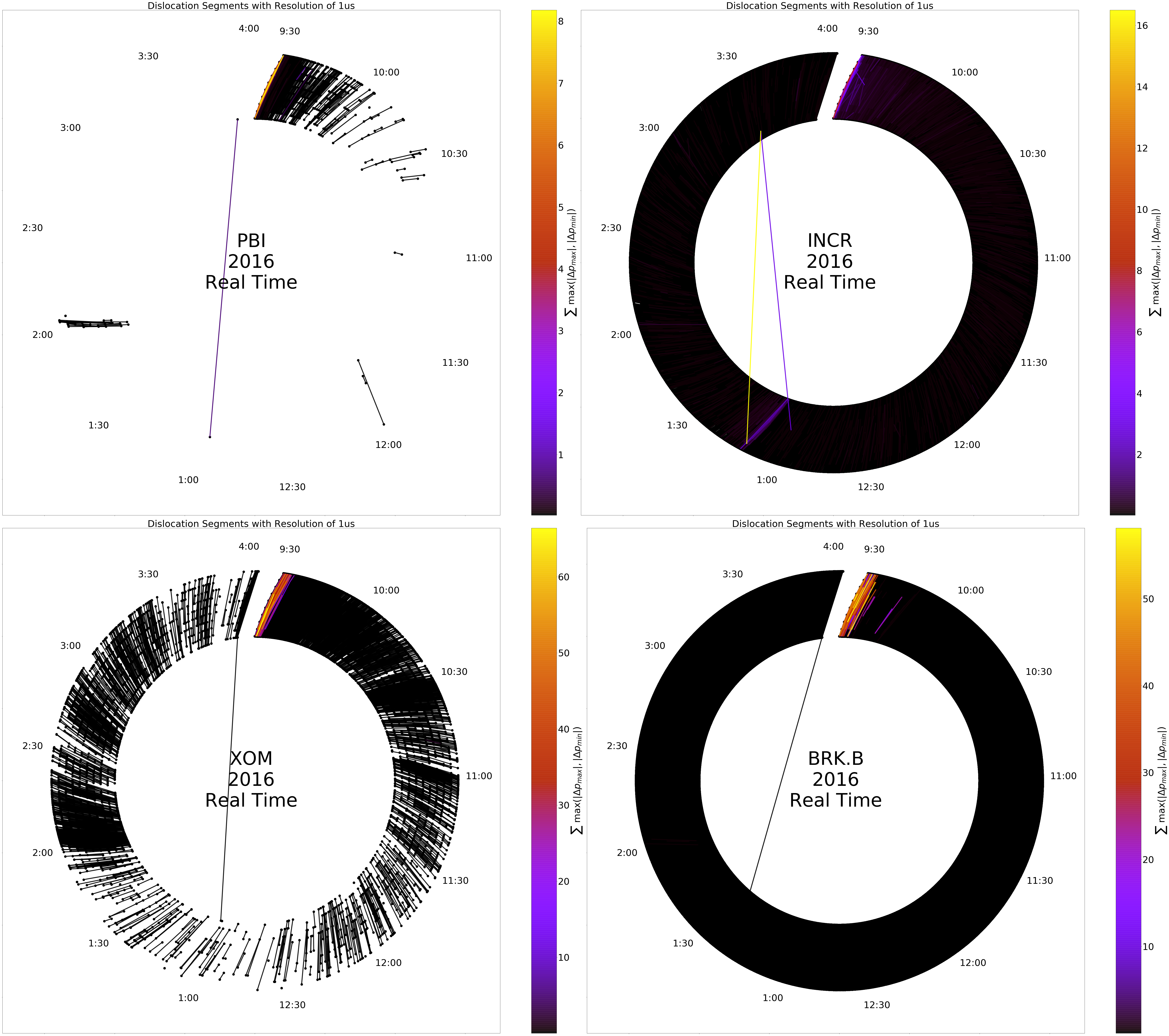}
    \caption{
    Dislocation segments (DS) for stock pairs (similar MC) aggregated over a year (modulo day).
    PBI (paired with INCR) is the smallest common stock by MC under consideration that remained in the S\&P 500 for all of 2016.
    BRK.B (paired with XOM) is the only mega cap in the RexSP.
    We see that DSs appear to be more concentrated for S\&P 500 constituents (left) with spikes occurring at the beginning of the trading day and at 2:00 pm.
    Additionally, we note that DSs appear to a smaller magnitude for S\&P 500 constituents.
    }
    \label{fig:circles}
\end{figure}

\paragraph{S\&P 500 Inclusion Effect: Dislocations}\label{para:snp_effect_dislocations}
As a visual aid to these results, we have included circle plots, as introduced in~\cite{tivnan2020fragmentation}, to demonstrate the non-uniform distribution of DSs that can occur.
Figure~\ref{fig:circles} shows these circle plots for two common stock pairs ((PBI, INCR), (BRK.B, XOM)) on the edges of our indices.
The first pair is the smallest common stock in the S\&P 500$\prime$ by MC that remained in the S\&P 500 for the entire calender year and the closest component by MC in the RexSP, PBI and INCR respectively.
The second pair is the only mega cap in the RexSP and the closest component by MC in the S\&P 500$\prime$ that remained in the S\&P 500 for the entire calender year, BRK.B and XOM respectively.
We note that BRK.A is not included in the Russell 3000~\citep{russmethods} and that XOM is additionally included in the DOW\@.
These common stock pairs underscore the difference in behavior between constituents of the S\&P 500 and those not included in the most worlds most widely tracked equity index.
\medskip

Figure~\ref{fig:circles} displays the DSs for the above-mentioned common stocks aggregated over a year (modulo day).
We see that DSs appear to be more concentrated for S\&P 500 constituents with spikes occurring at the beginning of the trading day and at 2:00 pm.
Additionally, DSs for S\&P 500 constituents tend to have smaller magnitudes, relative to Russell 3000 constituents.
We provide circle plots for many more securities on our webpage~\citep{compfi_page}.

\subsection{Market capitalization}\label{subsec:market-cap}

Further evidence for scaling behavior arises from analysis of MC\@.
Tables~\ref{tab:sectors} and~\ref{tab:market_cap_classes} display MC statistics broken down by industry sector and categorical size, e.g., micro-cap, mega-cap, etc.
MC is significantly positively correlated with ROC\@.
Tables~\ref{tab:roc-market-cap} -~\ref{tab:roc-market-cap-marginal-quad} display results from ordinary least squares regressions predicting ROC using MC and other predictors\@.
A linear fit predicting $\log_{10}$ \textit{ROC} from $\log_{10}$ \textit{MC}, $\log_{10}$ \textit{total trades}, and $\log_{10}$ \textit{differing trades} gives $R^2 \simeq 0.908$.
A positive coefficient relates $\log_{10}$ \textit{ROC} to $\log_{10}$ \textit{MC}, indicating that higher MC is associated with higher ROC\@.
A similar regression is computed including quadratic terms in $\log_{10}$ MC, which has a significant, but weak, negative association with ROC\@.
Similar relationships hold for both the linear and quadratic models when the dependent variable is instead chosen to be total or differing trades.
\medskip

\noindent
Though behavior of ROC as a function of MC is generally similar when equities are stratified by sector, some sectors display lower average levels of ROC, differing trades, or total trades when MC is held constant.
Equities classified as being in the financial sector generally have a smaller amount of ROC, while equities classified as being in the energy sector exhibit a higher amount of ROC on average.
However, there is no clear general trend linking sectors to MC or to ROC\@.

\subsection{Realized opportunity cost}\label{subsec:roc}
As expected with an increase in the number of analyzed equities from 30 to more than 2900, the amount of ROC rose substantially from the quantity reported in~\cite{tivnan2020fragmentation}, from \$160M to \$2.05B USD\@.
ROC clearly displays sublinear scaling with the number of studied equities;
we do not observe a thousandfold increase in the amount of ROC with a thousandfold increase in the number of equities.
The information advantage afforded traders with access to direct feed information is not uniform;
though a vast majority of the ROC (\$1.91 B) favored the direct feeds in this way, a non-negligible amount of ROC (\$137 M) did favor the SIP feeds.
Approximately a quarter (23.71\%) of all trades observed occurred during a dislocation.
The fraction of ``differing traded value''---the nominal market value of all differing trades---was slightly higher (25.25\%) than the fraction of all trades that were differing trades.
The ratio between these two values (25.25\% / 23.71\% = 1.0651) shows that the average differing trade moves approximately 6.51\% more value than the average trade.
This indicates a qualitative shift in trading behavior during dislocations.
\medskip

\noindent
Securities in the SPexDow account for a median of 2,006,091 differing trades per day, in contrast to the 309,158 in the Dow 30 or 1,921,121 in the RexSP\@.
The median differing traded value per day in the SPexDow was also the highest among the three categories, totaling approximately \$14.07T versus the RexSP's total of \$6.7T and the Dow's total of \$3.27T\@.
ROC per share differed across the three categories, with median ROC per share per day of 1.1\cent, 1.5\cent, 2.1\cent for the Dow, SPexDow, and RexSP respectively.
ROC per share tends to increase as MC decreases, with lowest ROC per share occurring in the Dow and highest ROC per share occurring in the RexSP\@.
Median total ROC per day on the Dow amounted to \$514.8K, while median total ROC per day on the SPexDow totaled \$3.384M and on the RexSP amounted to \$3.564M\@.
Summary statistics for distributions of ROC for each mutually-exclusive market category are given in Table~\ref{tab:purse-eyecatchers}.
\medskip

\begin{table}
    \tbl{
    Summary statistics for select common stock pairs.
    BRK.B (paired with XOM) is the only mega cap in the RexSP.
    PBI (paired with INCR) is the smallest common stock by MC under consideration that remained in the S\&P 500 for all of 2016.
    Note that those in the S\&P 500 (green) have a much higher trading volume and ROC then their similarly capitalized counterparts.
    }
    {\begin{tabular}{lrrrrrr}
        \toprule
        \tabhead{Ticker}{MC (\$)}{ROC (\$)}{Trades}{Diff. Trades}{Traded Value (\$)}{Diff. Traded Value (\$)}
        \midrule
        BRK.B & 401,644,421,120 & 2,278,835.98 & 5,120,595 & 1,544,050 & 70,435,832,686.71 & 24,162,015,573.13\\
        \rowcolor{ForestGreen!10}
        XOM & 374,280,552,448 & 8,846,416.18 & 16,146,652 & 4,479,209 & 169,057,336,872.77 & 47,541,675,580.93\\
        \rowcolor{ForestGreen!10}
        PBI & 2,821,674,240 & 726,596.69 & 2,360,470 & 488,092 & 5,766,285,837.56 & 1,257,265,907.34\\
        INCR & 2,820,235,520 & 487,049.13 & 904,613 & 243,855 & 3,989,174,661.59 & 1,016,834,174.82\\
        \bottomrule
    \end{tabular}}
    \label{tab:edge_roc}
\end{table}

\begin{table}
    \centering
    \sisetup{
    table-number-alignment = left,
    table-figures-integer  = 11,
    table-figures-decimal  = 2,
    }
    \tbl{
    Comparison of the smallest ten common stocks that remained in the S\&P 500 for all of 2016 (green) and the ten RexSP common stocks with the closest MC.
    Rows marked with $\dagger$ have significantly (two-sided t-test, p $<$ 0.05) higher values for common stocks in the S\&P 500.
    We note that common stocks in the S\&P 500 have nearly three times the trading activity and ROC than their similarly capitalized counterparts.
    }
    {\begin{tabular}{l S @{\ $\pm$ \ } S}
        \toprule
        \multicolumn{1}{c}{Stat.} & \multicolumn{1}{c}{Mean} & \multicolumn{1}{c}{Std.}\\
        \midrule
        \multirow{2}{*}{MC (\$)} & \greencell{3695890099.20} & \greencell{464930329.63}\\ & 3696678400.00 & 465021263.08\\[1ex]
        \multirow{2}{*}{ROC (\$)$^\dagger$} & \greencell{1530766.70} & \greencell{1212566.32}\\ & 573704.19 & 454901.87 \\[1ex]
        \multirow{2}{*}{Trades$^\dagger$} & \greencell{3757345.30} & \greencell{2579005.78}\\ & 1340988.30 & 1099357.71 \\[1ex]
        \multirow{2}{*}{Diff. Trades$^\dagger$} & \greencell{848648.80} & \greencell{568393.92}\\ & 318163.00 & 222699.69 \\[1ex]
        \multirow{2}{*}{Traded Value (\$)$^\dagger$} & \greencell{11966521828.32} & \greencell{6995211619.34}\\ & 4281159071.68 & 2466969453.45 \\[1ex]
        \multirow{2}{*}{Diff. Traded Value (\$)$^\dagger$} & \greencell{2930334696.21} & \greencell{1746456767.92} \\ & 1071563501.93 & 551246762.12 \\
        \bottomrule
    \end{tabular}}
    \label{tab:edge_exp}
\end{table}

\begin{figure}
    \centering
    \includegraphics[width=0.8\textwidth]{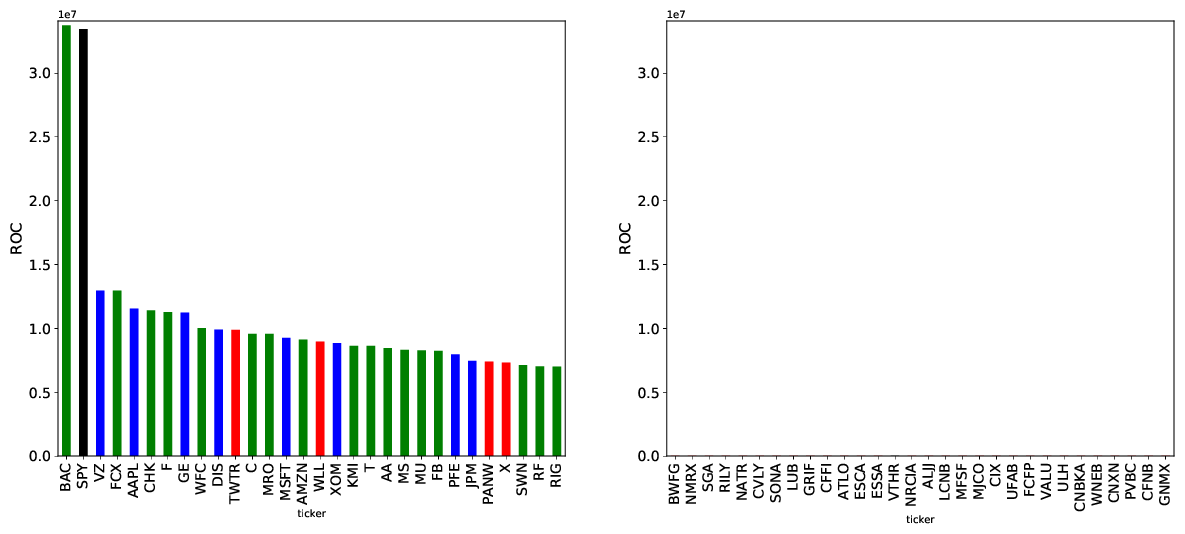}
    \caption{
    ROC by ticker (\$) for the top 30 (left panel) and bottom 30 (right panel) of all securities under study, ranked by ROC.
    Constituents of the Dow 30 are shown in blue, constituents of the S\&P 500 (excluding the Dow 30) are shown in green, constituents of the Russell 3000 (excluding the S\&P 500) are shown in red, and ETFs are shown in black.
    }
    \label{fig:all-top30-bottom30}
\end{figure}

\begin{figure}
    \centering
    \includegraphics[width=0.8\textwidth]{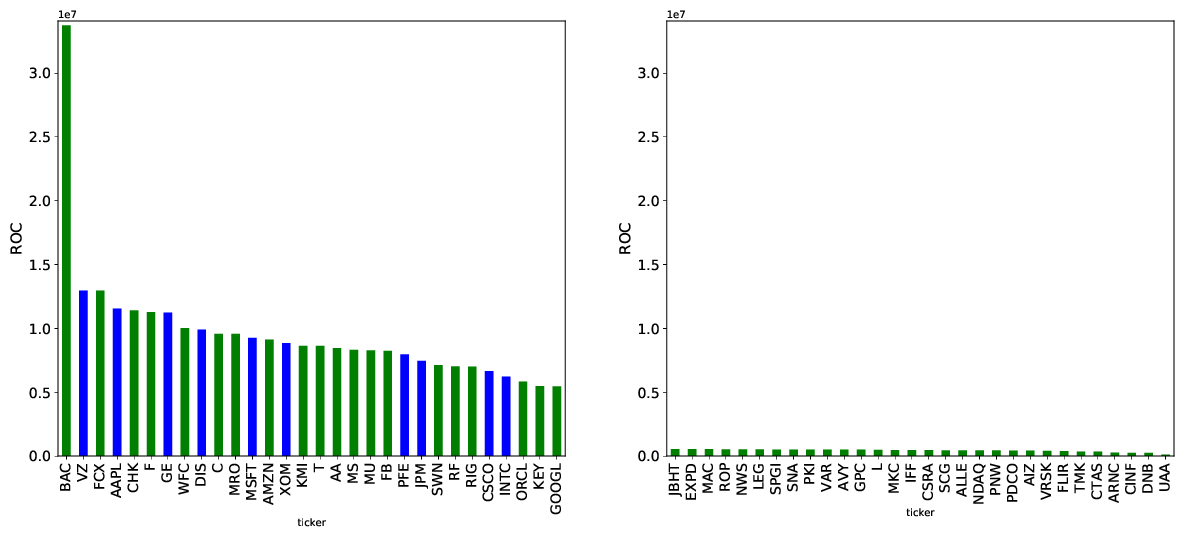}
    \caption{
    ROC by ticker (\$) for the top 30 (left panel) and bottom 30 (right panel) of S\&P 500 securities, ranked by ROC.
    Constituents of the Dow 30 are shown in blue, while those belonging to the S\&P 500 (excluding the Dow 30) are shown in green.
    }
    \label{fig:sp500-top30-bottom30}
\end{figure}

\begin{figure}
    \centering
    \includegraphics[width=0.8\textwidth]{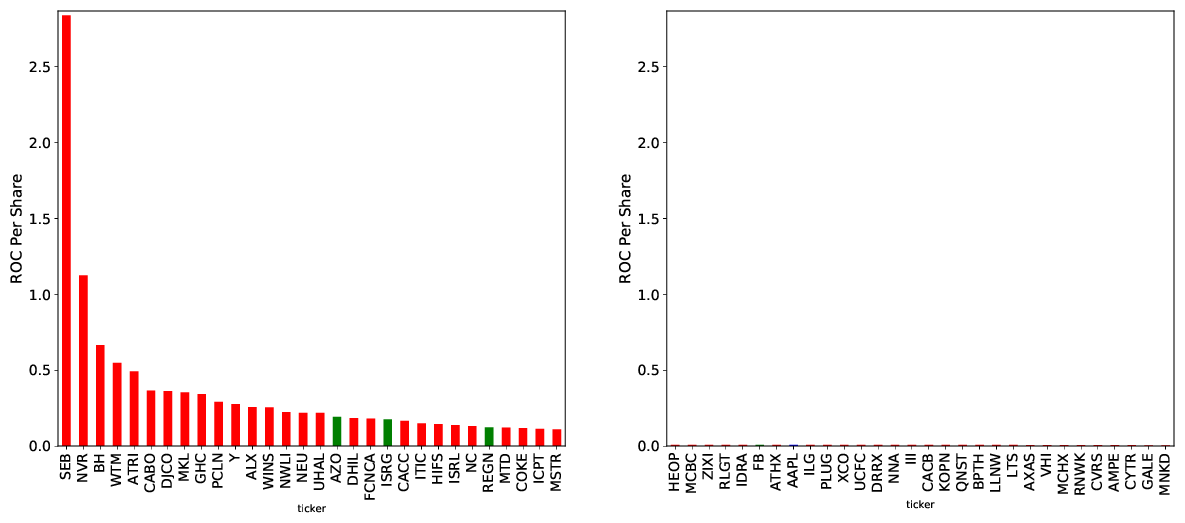}
    \caption{
    ROC per share (\$ / share) by ticker for the top 30 (left panel) and bottom 30 (right panel) of all securities under study, ranked by ROC.
    Constituents of the Dow 30 are shown in blue, constituents of the S\&P 500 (excluding the Dow 30) are shown in green, and constituents of the Russell 3000 (excluding the S\&P 500) are shown in red.
    }
    \label{fig:all-pershare-top30-bottom30}
\end{figure}

\noindent
It is interesting to consider the distribution of both total ROC and ROC per share by both equity and mutually-exclusive market category.
Figure~\ref{fig:all-top30-bottom30} displays ROC of the top 30 and bottom 30 of all securities under study when ranked by ROC\@.
Included in this figure for comparison is the exchange-traded fund SPY, an ETF that tracks the S\&P 500.
Selected ETFs are also treated separately in Section~\ref{subsec:etfs}.
It is notable that the equity with the largest ROC, Bank of America (BAC), has more than twice the ROC of the equity with the second-largest amount of ROC, Verizon (VZ).
Though not an equity and not included in the rest of this study, it is also notable that SPY, one of the most heavily traded securities on the NMS along with BAC, is close to BAC in ROC\@.
Of the top 30 securities with the most ROC, eight of the 30 are Dow 30 equities.
Only four out of 30 are RexSP equities, while the other 17 non-ETF securities are SPexDow equities.
One may attribute this to MC, though we note the S\&P 500 is not the largest 500 U.S.\ companies~\ref{sec:indices}.
In fact, there are 612 RexSP constituents with a MC greater than PBI, a common stock at the bottom of the S\&P 500.
This includes 67 large and mega cap common stocks.
Since the S\&P 500 appears to be the primary driver of ROC across all equities (c.f.\ below), we find the top 30 and bottom 30 S\&P 500 securities ranked by ROC, including Dow 30 securities, and plot their ROC in Figure~\ref{fig:sp500-top30-bottom30}.
Even in this subset, only 10 of the top 30 equities are Dow 30 securities.
However, when the unit of analysis changes to ROC per share, as in Figure~\ref{fig:all-pershare-top30-bottom30}, we find that RexSP equities fill 27 out of 30 top ranks, which corresponds with the aggregated statistics reported in Table~\ref{tab:purse-eyecatchers}.
Additionally, we revisit our common stock pairs from~\ref{para:snp_effect_dislocations} to take a closer look at common stocks barely inside and outside of the S\&P 500.
We see that the common stocks in the S\&P 500 have a much higher trading volume and ROC then their similarly capitalized counterparts~\ref{tab:edge_roc}.
To see if this trend holds we expand our set to the ten smallest common stocks that remained in the S\&P 500 for all of 2016 and the ten RexSP common stocks with the closest MC\@.
None of the ten RexSP members spent any time in the S\&P 500 during 2016.
We find the trend holds with members of the S\&P 500 having nearly three times the trading activity and ROC than their similarly capitalized counterparts~\ref{tab:edge_roc}.
\medskip

\begin{figure}
    \centering
    \includegraphics[width=0.8\textwidth]{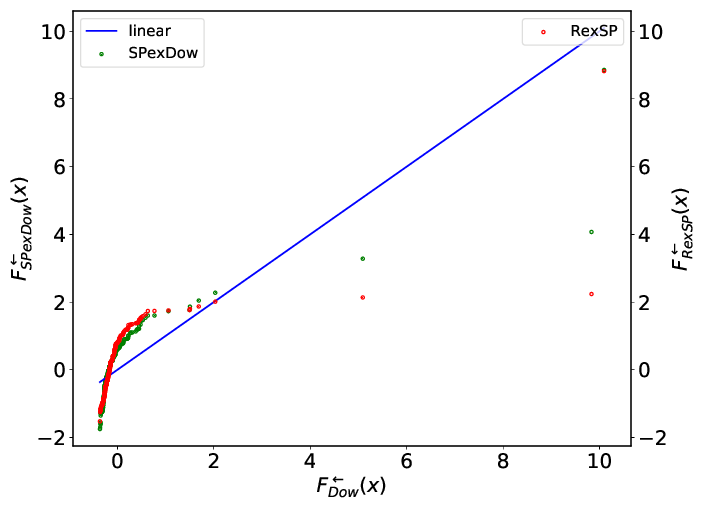}
    \caption{
    Empirical quantile-quantile (QQ) plot for the normalized ROC per share processes.
    It is clear that the distribution of the SPexDow and RexSP processes are similar, and both are markedly different from the Dow process (blue line).
    }
    \label{fig:roc-qq-plot}
\end{figure}

\noindent
Since there appear to be differences between the (stationary) summary statistics of the mutually-exclusive market categories, it is reasonable that there may be significant differences between the ROC statistics considered as time-dependent stochastic processes and simply considered as random variables decoupled from time.
Within each category, the ROC was computed for all equities in that category for each day.
Each ROC series is then normalized as $r_i \mapsto \frac{r_i - \langle r_i \rangle}{\sqrt{\text{Var}(r_i)}}$, which allows direct comparison of the series.
Figure~\ref{fig:roc-qq-plot} displays a quantile-quantile plot of the Dow, SPexDow, and RexSP ROC distributions.
The Dow distribution is plotted as linear and the other two distributions are compared with it.
It is immediately obvious that the left tails of the SPexDow and RexSP distributions are heavier than that of the Dow;
this also appears to be the case for the right tails of the distributions, but there is little sampling in this region and so no conclusion can be drawn.
This similarity of the SPexDow and RexSP distributions is also striking;
when normalized they appear almost identical.
\medskip

\begin{figure}
    \centering
    \includegraphics[width=0.8\textwidth]{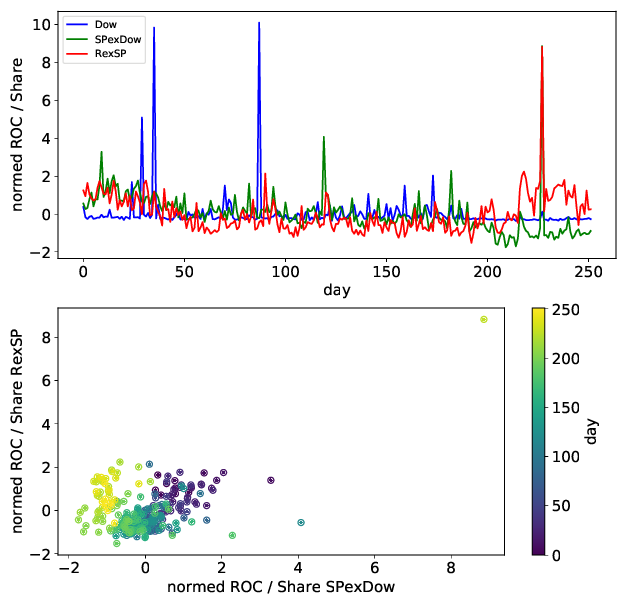}
    \caption{
    Normalized ROC per share processes.
    There is one observation per day for a total of 252 observations in the process.
    These processes are anti-autocorrelated (Dow DFA exponent $\alpha = 0.434$, SPexDow DFA exponent $\alpha = 0.226$, RexSP DFA exponent $\alpha = 0.301$) and exhibit rare large values.
    The lower panel provides evidence for nonlinear cross-correlation between the SPexDow and RexSP ROC per share processes.
    }
    \label{fig:roc-sample-paths}
\end{figure}

\noindent
Figure~\ref{fig:roc-sample-paths} displays the time-dependent sample paths of ROC sampled at daily resolution.
These processes are anti-autocorrelated---they display mean reversion---as evidenced by their detrended fluctuation analysis (DFA)~\citep{peng1994mosaic} exponents of $\alpha_{\text{Dow}} = 0.438$, $\alpha_{\text{SPexDow}} = 0.242$, and $\alpha_{\text{RexSP}} = 0.235$.
All series exhibit rare large values from time to time, with the Dow ROC series exhibiting the largest rare values relative to its mean fluctuations and the SPexDow series exhibiting the smallest.
We also note that, in accordance with the QQ plot of the time-decoupled distributions above, the DFA exponents of the SPexDow and RexSP---and thus their corresponding dynamical behavior---are closer than they are to the Dow DFA exponent.
\medskip

\begin{figure}
    \centering
    \includegraphics[width=.8\textwidth]{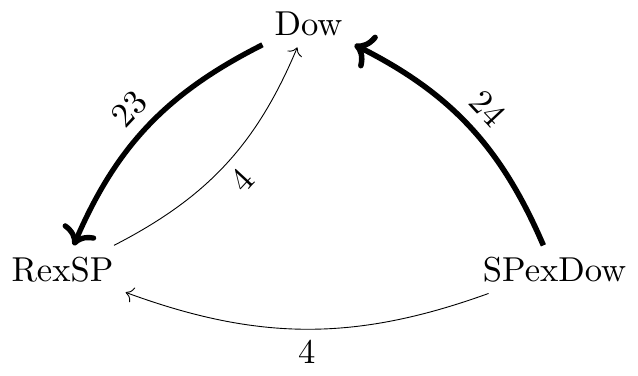}
    \caption{
    Network of relationships between mutually-exclusive market categories implied by results of four Granger causality tests.
    The direction of the edges gives the direction of the Granger-causal relationship, while the weight on the edge is the total number of lags for which the relationship was significant at the $p=0.05/N_{\text{lags}}$ level (the conservative Bonferroni correction).
    The maximum number of lags was chosen to be $N_{\text{lags}} = 40$.
    Thickness of the edge is proportional to edge weight and is plotted for emphasis in visualization.
    Details about which lags were associated with significant Granger causality can be found in Table~\ref{tab:granger-lags}.
    }
    \label{fig:roc-granger-cause}
\end{figure}

\noindent
A review of the above results points to the SPexDow as being the ``dominant" mutually-exclusive market category in some sense:
it accounts for a plurality of differing trades, differing traded value, and total ROC, while also having a DFA exponent lower than that of the Dow and close in value to that of the RexSP, meaning that its time-series of ROC is strongly mean-reverting.
The amalgamation of these facts can be interpreted as evidence that the SPexDow ROC time series is possibly least likely to be influenced by the other series of ROC\@.
To test this hypothesis, we conduct a number of Granger causality tests on the time series of ROC\@.
Granger causality is the notion that past values of one time series may be useful in predicting current and future values of another time series~\citep{granger1969investigating}.
A maximum lag of 40 days was set and four tests were calculated pairwise between each of the three mutually-exclusive market categories: sum of squared residuals $\chi^2$-test, a likelihood ratio test, sum of squared residuals $F$-test, and a Wald test.
We consider there to be a significant Granger causality between series when all four tests indicate significant Granger causality at the $p = 0.05 / N_{\text{lags}}$ confidence level.
The correction for multiple comparisons is done using the most conservative estimate, the Bonferroni correction, to minimize the probability of Type I error~\citep{bonferroni1936teoria}.
Figure~\ref{fig:roc-granger-cause} displays the results of these tests graphically as a directed network.
The direction of edges denotes the direction of the Granger-causal relationship between the categories, while the weights on the edges denote the total number of lags for which the relationship was significant.
The SPexDow is shown to significantly influence both the Dow and RexSP while not being significantly influenced by either category;
this provides strong evidence to support our above hypothesis.
We note that the SPY tracks the S\&P 500, is one of the most heavily-traded securities, and has the second-highest amount of ROC of the securities under study here.
The SPY's price dynamics and ROC may thus have a material effect on the relationships between the S\&P 500's ROC and those of the other market categories, providing a partial confounding effect to the Granger-causal relationship determined here;
there may be a mutually-causal relationship between the real S\&P 500 and the ETF that tracks it.
The RexSP and Dow have a mutually Granger-causal relationship, with the Dow exerting more influence on the RexSP than the other way around.
This finding corresponds with the ranking of categories on a total shares traded per number of equities basis;
this is not a surprising result.
We also find that the SPexDow exerts far less influence on the RexSP than does the Dow (four total lags for the SPexDow versus 23 total lags for the Dow), a fact for which we do not have a ready explanation.
\medskip

\noindent
Providing further evidence for the above hypothesis, we compute Pearson
correlations between pairs of mutually exclusive categories for both ROC
and ROC per share;
these results are displayed in Table~\ref{tab:roc-pearson-correlations}.

\begin{table}
    \tbl{
    Pearson correlation matrices of mutually-exclusive market categories.
    For each index subset a daily resolution time series is constructed for the given statistic over all stocks in the index subset.
    For the ROC series the ROC generated for each stock on a particular trading day is summed, while in the ROC per share case the values are averaged.
    The correlation coefficients are then calculated between pairs of time series in order to construct the tables above.
    The top table displays ROC correlations, while the bottom table displays ROC per share correlations.
    The ROC per share statistic normalizes the number of traded shares, allowing for a fair comparison between the more heavily traded stocks in the Dow 30 or S\&P 500 subset with the more lightly traded stocks in the Russell 3000 subset.
    }
    {\begin{tabular}{lrrr}
        \toprule
        \multicolumn{1}{l}{\textbf{ROC:}} & \multicolumn{1}{c}{Dow} & \multicolumn{1}{c}{SPexDow} & \multicolumn{1}{c}{RexSP}\\
        \midrule
        Dow     &  1.000000 &  0.451072 &  0.319018 \\
        SPexDow &  0.451072 &  1.000000 &  0.724903 \\
        RexSP   &  0.319018 &  0.724903 &  1.000000 \\
        \bottomrule
    \end{tabular}
    \qquad
    \begin{tabular}{lrrr}
        \toprule
        \multicolumn{1}{l}{\textbf{ROC / Share:}} & \multicolumn{1}{c}{Dow} & \multicolumn{1}{c}{SPexDow} & \multicolumn{1}{c}{RexSP}\\
        \midrule
        Dow     &  1.000000 & 0.103061 & -0.019662 \\
        SPexDow &  0.103067 & 1.000000 &  0.411443 \\
        RexSP   & -0.019662 & 0.411443 &  1.000000 \\
        \bottomrule
    \end{tabular}}
    \label{tab:roc-pearson-correlations}
\end{table}

ROC correlations are strongest between SPexDow and RexSP ($\rho = 0.72$) and SPexDow and Dow ($\rho = 0.45$), while the correlation between the RexSP and Dow is lower ($\rho = 0.31$).
ROC per share correlations are universally lower than those for ROC, but the correlations between SPexDow and RexSP ($\rho = 0.41$) and SPexDow and Dow ($\rho = 0.10$) are still higher than that between RexSP and Dow ($\rho = -0.01$), which is actually negative.
\medskip

\begin{figure}
    \centering
    \includegraphics[width=0.8\textwidth]{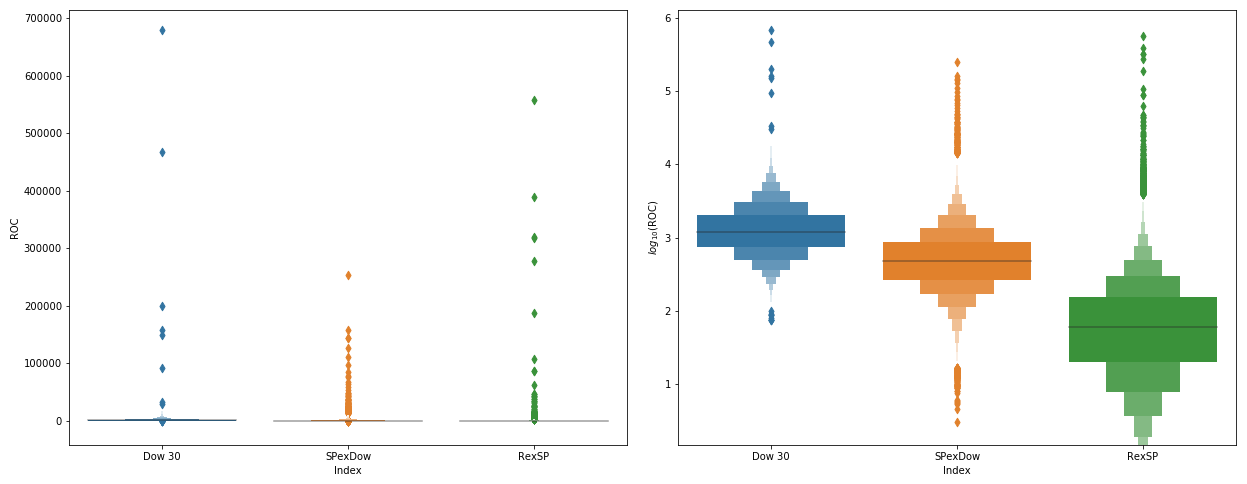}
    \caption{
    Distributions of mean ROC per day over the members of each mutually exclusive market category.
    Linear (left) and log 10 (right) vertical axis scaling are used to provide additional perspective.
    On average, members of the Dow experience more ROC than members of the SPexDow, which experience more ROC than the RexSP.
    These distributions are extremely heavy tailed, thus the use of log scaling, and feature a high degree of overlap.
    Thus there are members from each category that experience high ROC and low ROC.
    }
    \label{fig:mean-roc-bucket}
\end{figure}

\begin{figure}
    \centering
    \includegraphics[width=0.8\textwidth]{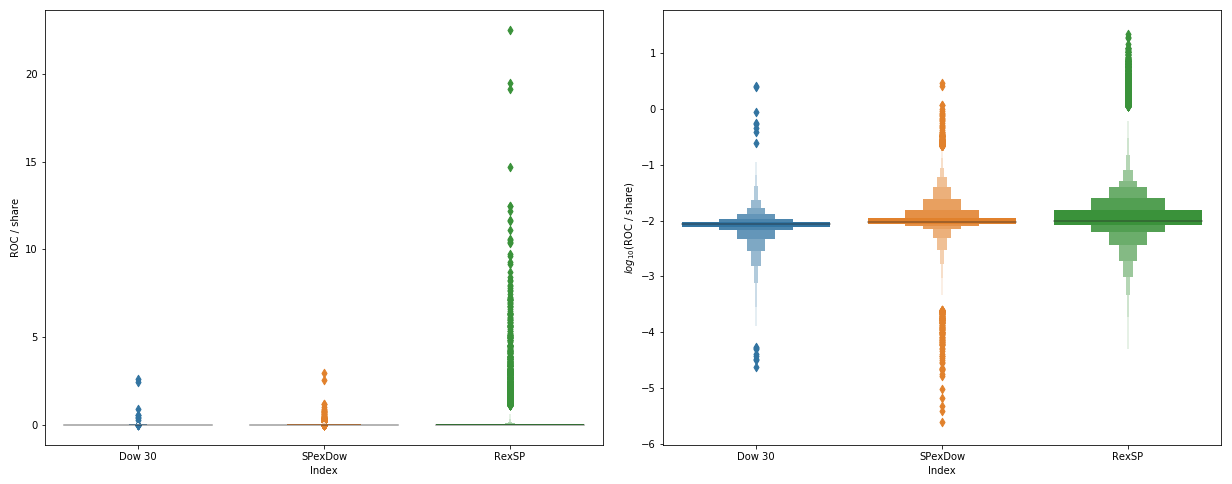}
    \caption{
    Distributions of mean ROC per share per day (\$ / day) over the members of each mutually exclusive market category.
    Linear (left) and log 10 (right) vertical axis scaling are used to provide additional perspective.
    On average, the members of the Dow experience the least ROC per share, followed by the SPexDow, followed by the RexSP.
    }
    \label{fig:mean-roc-per-share-bucket}
\end{figure}

\noindent
Figure~\ref{fig:mean-roc-bucket} displays the distributions of daily total ROC in 2016 by mutually-exclusive market category.
The panel with linear scaling highlights the extremely heavy-tailed nature of these distributions, while the log scaled panel provides a better comparison between the mutually-exclusive market categories.
On average, members of the Dow 30 experience the greatest daily ROC, followed by members of the SPexDow, followed by members of the RexSP\@.
It seems likely that the kurtosis of the theoretical distributions do not exist, implying tail exponent $\gamma < 4$ in the distribution $\Pr(X > x) \sim x^{-(\gamma -1)}$.
Table~\ref{tab:bucket-daily-roc-kurtosis-skew} displays the skew and kurtosis for each distribution.
If we examine the daily ROC per share in a similar manner, which is shown in Figure~\ref{fig:mean-roc-per-share-bucket}, we observe a reversal of the previous relationship.
Members of the Dow 30 have the least daily ROC per share, on average, and members of the RexSP have the most.
Though there is a slight trend, more ROC per share in less frequently traded stocks, the distributions of all three groups are nearly centered at 1\cent\ per share.
This corresponds well with our expectations based on the structure of the system and the distribution of DS magnitudes shown in Figure~\ref{fig:ds-rexsp-example}.

\subsection{ETFs}\label{subsec:etfs}
Exchange traded funds (ETFs) are securities that trade on the NMS and are designed to mimic as closely as possible a particular portfolio of other securities.
They are thus governed by the same price discovery mechanism as other securities that trade on the NMS, as opposed to the end-of-day price discovery mechanism to which mutual funds are subjected, but also allow investors to own a portion of potentially many underlying assets (or at least a simulacrum of such), similar to a mutual fund.
Here, we briefly remark on the similarities and differences between ETFs designed to track subsets of the market and those subsets of the market themselves.
We calculate statistics on the DSs and ROC associated to ETFs from three firms (Vanguard, iShares, Russell) for three indices (S\&P 500, Russell 300, Russell 2000), for a total of nine ETFs (SPY, VOO, IVV, THRK, VTHR, IWV, TWOK, VTWO, IWN).
The Russell 2000 is comprised of the smallest 2000 firms in the Russell 3000 by MC\@.
The S\&P 500 and Russell 3000 were selected as measures of overall market activity while the Russell 2000 was selected to isolate dynamics among ETFs that track smaller equities.
\medskip

\noindent
Table~\ref{tab:etfs-eyecatchers} summarizes ROC statistics for the ETFs under study.
The fraction of differing trades and differing traded value are lower than for any of the indexes as a whole;
in fact, the ratio of the fraction of differing traded value to the fraction of differing trades is less than one.
Total ROC incurred from trades in ETFs studied here totaled over \$38 million in calendar year 2016.
This statistic provides some evidence to suggest that ETFs
have their own endogenous statistical behavior that
differs from the behavior of the assets from which they are derived.

\section{Conclusion}\label{sec:conclusion}
In sum, we have demonstrated that the existence of DSs and ROC is not restricted to Dow 30 securities.
Furthermore, we have established that these microstructure quantities occur with non-negligible frequency and size;
we show that total ROC in Russell 3000 securities was well in excess of \$2 billion USD during 2016.
While consistent with the two comprehensive studies of the modern U.S.\ market~\citep{wah2016prevalent, aquilina2020quantifying}, our ROC calculations provide the first empirical evidence explaining how traders might profitably exploit market dislocations, despite paying up to \$2.0B USD annually for order flow~\citep{wursthorn2020robinhood}.
\medskip

\noindent
Compounding these results, we provide strong statistical evidence that the S\&P 500 excluding Dow 30 securities, to which we refer as the SPexDow, is the primary driver of ROC among the three mutually exclusive categories of equities (Dow 30, SPexDow, and Russell 3000 excluding S\&P 500 securities, or the RexSP).
\medskip

\noindent
Compounding the above results, we find that structure in the distributions of DS start times and duration persist across the entire Russell 3000, indicating some broader microstructure-based proximate cause of this structure.
Distributions of DS duration exhibit a large peak between $10^{-4}$ and $10^{-3}$ seconds (100 microseconds to one millisecond), but also exhibit a second smaller, yet distinct, peak near one second.
This separation of timescales in the distribution provide evidence for the existence of at least two distinct proximate causes of DS.
Distributions of DS start times display even more intricate structure, with large peaks at the beginning and end of the trading day, self-similarity on the half-hour and ten-minute timescales, and a large spike at 2:00 P\@..
\medskip

\noindent
ROC was highest among SPexDow securities, but ROC per share was highest among RexSP securities, which were also the most lightly-traded securities.
All time series of ROC exhibit behavior of anti-autocorrelation, meaning that they are mean-reverting.
ROC in the SPexDow Granger-cause ROC in the other market categories, but the converse is not true;
while the Dow Granger-causes the RexSP, the RexSP only weakly Granger-causes the Dow and does not have any effect on the SPexDow.
\medskip

\noindent
Taken together, these results paint the picture of a NMS the physical structure of which generates effects that are persistent across size of equity and exchange.
Amplifying these persistent effects is the apparent central role of the SPexDow;
in number of DSs, amount of ROC, spectral properties of ROC time series, and Granger-causal relationships, the story emerges of the SPexDow's characteristics being generated by largely-endogenous factors and subsequently influencing the characteristics of the Dow and RexSP\@.
Future work could explore in more depth the extent to which microstructure effects arising first in the SPexDow then spread to other mutually exclusive market categories and propagate through time.
This work could also explore the evolutionary dynamics of the modern NMS from its birth following the financial crisis of 2007/8 to the present day.
The NMS may not have remained static, with a constant number of market centers and a stationary distribution of market agents and trading strategies, but rather may have experienced fluctuations in the number of exchanges, in the regulatory environment, and in strategy profiles of trading agents.
Such an analysis could pave the way for better informed modelling efforts and the advancement of market theory.

\section*{Acknowledgements}
The authors gratefully acknowledge helpful discussions with Anshul Anand, Yosry Barsoum, Lashon Booker, David Bringle, Eric Budish, Peter Carrigan, Gary Comparetto, Peter Sheridan Dodds, Bryanna Dienta, Jordan Feidler, Andre Frank, Bill Gibson, Frank Hatheway, Michelle Herd, Emily Hiner, Chuck Howell, Eric Hunsader, Robert Jackson, Neil Johnson, Matthew Koehler, Blake LeBaron, Connie Lewis, Matthew McMahon, Wade Shen, David Slater, Jonathan Smith, Paul Soldate, Brendan Tivnan, Jason Veneman, Elaine Wah, Sanith Wijesinghe, and Thomas Wilk.

\section*{Funding}
All opinions and remaining errors are the sole responsibility of the authors and do not reflect the opinions nor perspectives of their affiliated institutions nor that of the funding agencies.
D.R.D., C.M.V.O., J.H.R., and B.F.T.\ were supported by DARPA award \#W56KGU-17-C-0010.
C.M.D.\ was supported by NSF grant \#1447634.
C.M.D.\ was supported by a gift from Massachusetts Mutual Life Insurance Company.
The authors declare no conflicts of interest.
The views, opinions and/or findings expressed are those of the authors and should not be interpreted as representing the official views or policies of the Department of Defense or the U.S.\ Government.

\bibliographystyle{rQUF}
\bibliography{bmc_article}

\appendices
\section{Tables}\label{sec:tables}

\begin{table}[h]
    \tbl{
    Mean of dislocation segment summary statistics taken across the 30 members of the Dow.
    $545 \mu s$ is used for duration conditioning and \$$0.01$ is used for magnitude conditioning.
    }
    {\begin{tabular}{llrrr}
        \toprule
        \tabhead{Conditioned}{}{min magnitude (\$)}{max magnitude (\$)}{duration (s)}
        \midrule
        \multirow{8}{*}{None}        & count & 4,011,848.7333 &                & \\
        & mean  &         0.0110 &         0.0136 &         0.075413 \\
        & std   &         0.0391 &         0.2725 &         5.829465 \\
        & min   &         0.0100 &         0.0100 &      $<$0.000001 \\
        & 25\%  &         0.0100 &         0.0100 &         0.000248 \\
        & 50\%  &         0.0100 &         0.0100 &         0.000669 \\
        & 75\%  &         0.0100 &         0.0103 &         0.001253 \\
        & max   &        44.6933 &       279.2057 &     8,408.931478 \\
        \midrule
        \multirow{8}{*}{Duration}    & count & 2,169,106.5333 &                &\\
        & mean  &         0.0108 &         0.0149 &         0.132779 \\
        & std   &         0.0436 &         0.3548 &         7.645375 \\
        & min   &         0.0100 &         0.0100 &         0.000546 \\
        & 25\%  &         0.0100 &         0.0100 &         0.000783 \\
        & 50\%  &         0.0100 &         0.0100 &         0.001129 \\
        & 75\%  &         0.0100 &         0.0107 &         0.002654 \\
        & max   &        43.4150 &       279.1987 &     8,408.931478 \\
        \midrule
        \multirow{8}{*}{\shortstack{Duration \&\\ Magnitude}} & count &   95,757.8000 &               &\\
        & mean  &        0.0427 &        0.2370 &      0.955731 \\
        & std   &        0.3355 &        1.6130 &     48.214785 \\
        & min   &        0.0200 &        0.0200 &      0.000546 \\
        & 25\%  &        0.0200 &        0.0200 &      0.000698 \\
        & 50\%  &        0.0200 &        0.0227 &      0.001073 \\
        & 75\%  &        0.0307 &        0.0433 &      0.003552 \\
        & max   &       43.4150 &      114.3480 &  7,186.866464 \\
        \bottomrule
    \end{tabular}}
    \label{tab:ds-dow}
\end{table}

\begin{table}
    \tbl{
    Mean of dislocation segment summary statistics taken across 446 members of the SPexDow.
    $545 \mu s$ is used for duration conditioning and \$$0.01$ is used for magnitude conditioning.
    }{
    \begin{tabular}{llrrr}
        \toprule
        \tabhead{Conditioned}{}{min magnitude (\$)}{max magnitude (\$)}{duration (s)}
        \midrule
        \multirow{8}{*}{None}       & count & 2,525,082.0448 &                &\\
        & mean  &         0.0135 &         0.0168 &         0.252981 \\
        & std   &         0.2801 &         0.3996 &         9.325161 \\
        & min   &         0.0100 &         0.0100 &      $<$0.000001 \\
        & 25\%  &         0.0100 &         0.0100 &         0.000227 \\
        & 50\%  &         0.0100 &         0.0101 &         0.000583 \\
        & 75\%  &         0.0115 &         0.0136 &         0.001085 \\
        & max   &       476.1177 &       522.6072 &     9,084.040084 \\
        \midrule
        \multirow{8}{*}{Duration}   & count & 1,189,460.6682 &                &\\
        & mean  &         0.0134 &         0.0185 &         0.555820 \\
        & std   &         0.4601 &         0.6076 &        13.029491 \\
        & min   &         0.0100 &         0.0100 &         0.000546 \\
        & 25\%  &         0.0100 &         0.0100 &         0.000754 \\
        & 50\%  &         0.0102 &         0.0107 &         0.001119 \\
        & 75\%  &         0.0117 &         0.0160 &         0.008169 \\
        & max   &       471.7331 &       515.4222 &     9,084.040084 \\
        \midrule
        \multirow{8}{*}{\shortstack{Duration \&\\ Magnitude}} & count &  114,770.0224 &               &\\
        & mean  &        0.0557 &        0.1249 &       1.591543 \\
        & std   &        1.9177 &        2.5050 &      54.064998 \\
        & min   &        0.0200 &        0.0200 &       0.000546 \\
        & 25\%  &        0.0202 &        0.0209 &       0.000717 \\
        & 50\%  &        0.0228 &        0.0346 &       0.001240 \\
        & 75\%  &        0.0375 &        0.0625 &       0.027820 \\
        & max   &      471.7331 &      506.9715 &   6,943.106256 \\
        \bottomrule
    \end{tabular}}
    \label{tab:ds-spexdow}
\end{table}

\begin{table}
    \tbl{
    Mean of dislocation segment summary statistics taken across the 2451 members of the RexSP.
    $545 \mu s$ is used for duration conditioning and \$$0.01$ is used for magnitude conditioning.
    }
    {\begin{tabular}{llrrr}
        \toprule
        \tabhead{Conditioned}{}{min magnitude (\$)}{max magnitude (\$)}{duration (s)}
        \midrule
        \multirow{8}{*}{None}        & count &  770,577.8246 &               &\\
        & mean  &        0.9734 &        1.1361 &       4.413179 \\
        & std   &       34.0534 &       37.7472 &      50.079342 \\
        & min   &        0.0100 &        0.0100 &    $<$0.000001 \\
        & 25\%  &        0.0116 &        0.0121 &       0.000245 \\
        & 50\%  &        0.0139 &        0.0149 &       0.001042 \\
        & 75\%  &        0.0225 &        0.0302 &       0.013774 \\
        & max   &    2,238.1205 &    2,514.9617 &   8,796.956807 \\
        \midrule
        \multirow{8}{*}{Duration}    & count &  287,399.7217 &               &\\
        & mean  &        1.2116 &        1.7162 &      12.749530 \\
        & std   &       37.6277 &       46.3599 &      83.465004 \\
        & min   &        0.0100 &        0.0100 &       0.000546 \\
        & 25\%  &        0.0110 &        0.0118 &       0.002065 \\
        & 50\%  &        0.0147 &        0.0188 &       0.072213 \\
        & 75\%  &        0.0263 &        0.0408 &       0.975526 \\
        & max   &    2,033.1633 &    2,302.4541 &   8,796.956807 \\
        \midrule
        \multirow{8}{*}{\shortstack{Duration \&\\ Magnitude}} & count &   45,062.3366 &               &\\
        & mean  &        2.1734 &        3.0486 &     13.154607 \\
        & std   &       53.2211 &       66.0958 &    112.101259 \\
        & min   &        0.0200 &        0.0200 &      0.000546 \\
        & 25\%  &        0.0239 &        0.0272 &      0.003933 \\
        & 50\%  &        0.0338 &        0.0449 &      0.053583 \\
        & 75\%  &        0.0611 &        0.0806 &      0.798791 \\
        & max   &    2,033.9931 &    2,295.6782 &  7,139.075345 \\
        \bottomrule
    \end{tabular}}
    \label{tab:ds-rexsp}
\end{table}

\begin{table}
    \tbl{
    Mean of dislocation segment summary statistics taken across the 9 ETFs under study.
    $545 \mu s$ is used for duration conditioning and \$$0.01$ is used for magnitude conditioning.
    }
    {\begin{tabular}{llrrr}
        \toprule
        \tabhead{Conditioned}{}{min magnitude (\$)}{max magnitude (\$)}{duration (s)}
        \midrule
        \multirow{8}{*}{None}       & count & 6,431,595.4444 &                &\\
        & mean  &         0.0216 &         0.0273 &         0.339145 \\
        & std   &         0.0856 &         0.1027 &        13.327128 \\
        & min   &         0.0100 &         0.0100 &      $<$0.000001 \\
        & 25\%  &         0.0100 &         0.0100 &         0.000284 \\
        & 50\%  &         0.0100 &         0.0100 &         0.000602 \\
        & 75\%  &         0.0122 &         0.0156 &         0.001175 \\
        & max   &         9.0956 &         9.3744 &     5,658.596041 \\
        \midrule
        \multirow{8}{*}{Duration}    & count & 3,674,884.7778 &                  &\\
        & mean  &         0.0223 &         0.0289 &         0.683211 \\
        & std   &         0.0859 &         0.1077 &        18.991011 \\
        & min   &         0.0100 &         0.0100 &         0.000546 \\
        & 25\%  &         0.0100 &         0.0100 &         0.000726 \\
        & 50\%  &         0.0100 &         0.0111 &         0.001064 \\
        & 75\%  &         0.0122 &         0.0167 &         0.002494 \\
        & max   &         6.3278 &         8.4556 &     5,658.596041 \\
        \midrule
        \multirow{8}{*}{\shortstack{Duration \&\\ Magnitude}} & count &  130,853.7778 &               &\\
        & mean  &        0.1707 &        0.1800 &       0.933693 \\
        & std   &        0.2804 &        0.2995 &      26.558084 \\
        & min   &        0.0200 &        0.0200 &       0.000546 \\
        & 25\%  &        0.0200 &        0.0200 &       0.000765 \\
        & 50\%  &        0.0344 &        0.0411 &       0.001213 \\
        & 75\%  &        0.1733 &        0.2933 &       0.005725 \\
        & max   &        6.3278 &        8.4311 &   5,005.870452 \\
        \bottomrule
    \end{tabular}}
    \label{tab:ds-etf}
\end{table}

\begin{table}
    \tbl{
    Summary statistics of realized opportunity cost (ROC) for various equity groups under study during 2016.
    }
    {\begin{tabular}{lllr}
        \toprule
        \multirow{10}{*}{Russ 3K$^{\prime}$} & 1  & Realized Opportunity Cost   &      \$2,013,458,668.87\\
        & 2  & SIP Opportunity Cost        &      \$1,876,048,519.06   \\
        & 3  & Direct Opportunity Cost     &        \$137,410,149.76   \\
        & 4  & Trades                      &        4,658,307,833      \\
        & 5  & Diff.\ Trades               &        1,105,201,803      \\
        & 6  & Traded Value                & \$24,352,760,600,270.47   \\
        & 7  & Diff.\ Traded Value         &  \$6,272,439,590,589.91   \\
        & 8  & Percent Diff.\ Trades       &                   23.73   \\
        & 9  & Percent Diff.\ Traded Value &                   25.76   \\
        & 10 & Ratio of 9 / 8              &                    1.0855 \\
        \midrule
        \multirow{10}{*}{RexSP} & 1 & Realized Opportunity Cost & \$948,743,328.62\\
        & 2 & SIP Opportunity Cost &      \$911,950,130.85 \\
        & 3 & Direct Opportunity Cost &   \$36,793,197.77 \\
        & 4 & Trades &                    2,093,415,072 \\
        & 5 & Diff.\ Trades &              482,055,297 \\
        & 6 & Traded Value &              \$6,669,357,410,332.23 \\
        & 7 & Diff.\ Traded Value &        \$1,705,272,719,045.67 \\
        & 8 & Percent Diff.\ Trades &        23.03 \\
        & 9 & Percent Diff.\ Traded Value &  25.57 \\
        & 10 & Ratio of 9 / 8 &          1.1104 \\
        \midrule
        \multirow{10}{*}{S\&P 500$^{\prime}$} & 1 & Realized Opportunity Cost & \$1,064,715,340.25\\
        & 2 & SIP Opportunity Cost &      \$964,098,388.26  \\
        & 3 & Direct Opportunity Cost &   \$100,616,951.99 \\
        & 4 & Trades &                    2,564,892,761 \\
        & 5 & Diff.\ Trades &              623,146,506 \\
        & 6 & Traded Value &              \$18,429,250,470,003.83 \\
        & 7 & Diff.\ Traded Value &        \$4,567,166,871,544.24 \\
        & 8 & Percent Diff.\ Trades &         24.30 \\
        & 9 & Percent Diff.\ Traded Value &   25.83 \\
        & 10 & Ratio of 9 / 8 &          1.0631 \\
        \midrule
        \multirow{10}{*}{SPexDow} & 1 & Realized Opportunity Cost & \$904,501,417.30\\
        & 2 & SIP Opportunity Cost &  \$842,017,261.86 \\
        & 3 & Direct Opportunity Cost &   \$62,484,155.44 \\
        & 4 & Trades &                    2,172,791,182 \\
        & 5 & Diff.\ Trades &               535,714,275 \\
        & 6 & Traded Value &              \$13,824,440,155,934.76 \\
        & 7 & Diff.\ Traded Value &        \$3,666,630,946,582.52 \\
        & 8 & Percent Diff.\ Trades &         24.66 \\
        & 9 & Percent Diff.\ Traded Value &   26.52 \\
        & 10 & Ratio of 9 / 8 &          1.0757 \\
        \midrule
        \multirow{10}{*}{Dow 30$^{\prime}$} & 1  & Realized Opportunity Cost   &       \$160,213,922.95\\
        & 2  & SIP Opportunity Cost        &       \$122,081,126.40   \\
        & 3  & Direct Opportunity Cost     &        \$38,132,796.55   \\
        & 4  & Trades                      &         392,101,579      \\
        & 5  & Diff.\ Trades               &          87,432,231      \\
        & 6  & Traded Value                & \$3,858,963,034,003.48   \\
        & 7  & Diff.\ Traded Value         &   \$900,535,924,961.72   \\
        & 8  & Percent Diff.\ Trades       &                  22.30   \\
        & 9  & Percent Diff.\ Traded Value &                  23.34   \\
        & 10 & Ratio of 9 / 8              &                   1.0465 \\
        \bottomrule
    \end{tabular}}
    \label{tab:roc-eyecatchers}
\end{table}

\begin{table}
    \tbl{
    Purse statistics for all stocks under study in 2016.
    The data used to construct this table is aggregated by date and stock, resulting in 720,991 data points that correspond with the 731,556 combinations of 252 trading days in 2016 and 2903 stocks under study.
    }
    {\begin{tabular}{lrrrrrr}
        \toprule
        \tabhead{}{Trades}{Traded Value (\$)}{Diff.\ Trades}{Diff.\ Traded Value (\$)}{ROC (\$)}{ROC/Share}
        \midrule
        count &   720,991 &          720,991 &       720,991 &             720,991 &      720,991 &    720,991 \\
        mean  &  6,460.98 &    33,776,788.61 &      1,532.89 &        8,699,747.42 &     2,792.63 &   0.020880 \\
        std   & 13,249.67 &   109,021,779.70 &      3,036.98 &       25,738,960.57 &    17,611.14 &   0.087810 \\
        min   &         0 &                0 &             0 &                   0 &            0 &          0 \\
        25\%  &       599 &     1,118,022.02 &           101 &          199,882.83 &     237.6100 &   0.009510 \\
        50\%  &     2,020 &     5,316,322.22 &           450 &        1,246,241.41 &     826.6000 &   0.011448 \\
        75\%  &     6,478 &    24,797,793.44 &         1,600 &        6,525,124.17 &     2,578.75 &   0.018836 \\
        max   &   517,270 & 8,280,915,338.59 &       103,885 &    1,596,912,962.05 & 6,798,041.07 &    19.3381 \\
        \bottomrule
    \end{tabular}}
    \label{tab:purse-all}
\end{table}

\begin{table}
    \tbl{
    Aggregated purse statistics for different groups of securities in 2016.
    Each section is composed of date aggregated data, resulting in 252 data points that correspond with the 252 trading days in 2016.
    }
    {\begin{tabular}{llrrrrrr}
        \toprule
        \tabhead{}{}{Trades}{Traded Value (\$)}{Diff.\ Trades}{Diff.\ Traded Value (\$)}{ROC (\$)}{ROC/Share}
        \midrule
        \multirow{7}{*}{Russ 3K$^{\prime}$}  & mean  & 18,485,348.54 &  96,637,938,889.96 &  4,385,721.44 &   24,890,633,295.99 &  7,989,915.35 &   0.023073 \\
        & std   &  3,705,825.95 &  17,507,577,514.36 &  1,222,558.47 &    5,929,581,247.64 &  2,363,234.20 &   0.003143 \\
        & min   &     7,045,815 &  41,324,500,835.46 &     1,197,040 &    8,277,978,080.59 &  2,717,631.16 &   0.018414 \\
        & 25\%  &    16,178,390 &  85,348,849,125.71 &     3,674,541 &   21,481,677,427.57 &  6,560,601.80 &   0.020888 \\
        & 50\%  & 17,837,416.50 &  94,176,286,443.74 &  4,257,438.50 &   24,165,074,815.55 &  7,524,560.38 &   0.022379 \\
        & 75\%  & 20,114,165.50 & 103,932,196,142.46 &  4,964,932.50 &   27,054,706,014.87 &  8,884,110.40 &   0.024693 \\
        & max   &    32,913,872 & 169,395,493,215.29 &     9,253,338 &   47,500,228,278.03 & 19,622,594.00 &   0.051371 \\
        \midrule
        \multirow{7}{*}{RexSP} & mean  &  8,307,202.67 &  26,465,704,009.25 &  1,912,917.85 &    6,766,955,234.31 &  3,764,854.48 &   0.022109 \\
        & std   &  1,370,512.88 &   3,786,979,882.64 &    473,884.96 &    1,299,054,438.46 &  1,048,372.83 &   0.002874 \\
        & min   &     3,183,224 &  11,363,776,182.38 &       487,500 &    2,268,729,995.29 &  1,436,093.46 &   0.017744 \\
        & 25\%  &  7,528,810.25 &  24,222,297,224.76 &  1,648,499.25 &    6,053,458,251.52 &  3,182,173.91 &   0.020092 \\
        & 50\%  &  8,175,352.50 &  26,166,834,634.22 &  1,921,121.50 &    6,779,433,456.68 &  3,564,482.05 &   0.021393 \\
        & 75\%  &  9,061,096.50 &  28,685,877,060.20 &  2,161,350.50 &    7,599,965,429.85 &  4,206,538.80 &   0.023737 \\
        & max   &    13,408,508 &  41,337,807,991.92 &     3,537,890 &   10,627,257,029.61 & 10,083,342.57 &   0.047415 \\
        \midrule
        \multirow{7}{*}{S\&P 500$^{\prime}$} & mean  & 10,178,145.88 &  70,172,234,880.71 &  2,472,803.60 &   18,123,678,061.68 &  4,225,060.87 &   0.014624 \\
        & std   &  2,406,751.15 &  14,303,150,882.94 &    775,201.38 &    4,760,162,875.50 &  1,531,548.30 &   0.002019 \\
        & min   &     3,862,591 &  29,960,724,653.08 &       709,540 &    5,941,906,620.96 &  1,281,537.70 &   0.011127 \\
        & 25\%  &  8,716,552.50 &  60,764,387,798.11 &  2,034,844.50 &   15,251,685,767.67 &  3,371,948.52 &   0.013502 \\
        & 50\%  &     9,684,039 &  67,776,548,100.32 &     2,310,806 &   17,479,288,594.91 &  3,918,496.70 &   0.014407 \\
        & 75\%  & 11,120,226.50 &  75,672,607,052.02 &  2,783,838.50 &   20,074,235,595.26 &  4,654,693.39 &   0.015434 \\
        & max   &    19,505,364 & 128,057,685,223.37 &     5,715,448 &   37,114,729,300.67 & 14,335,072.09 &   0.031484 \\
        \midrule
        \multirow{7}{*}{SPexDow}             & mean  &  8,622,187.23 &  54,858,889,507.68 &  2,125,850.30 &   14,550,122,803.90 &  3,589,291.34 &   0.014818 \\
        & std   &  1,960,102.37 &  10,686,728,768.81 &    632,025.23 &    3,571,347,460.11 &  1,119,395.15 &   0.002029 \\
        & min   &     3,283,385 &  23,296,053,599.93 &       619,976 &    4,906,051,591.25 &  1,136,332.05 &   0.011271 \\
        & 25\%  &  7,398,970.25 &  48,123,050,130.46 &  1,762,152.75 &   12,329,749,894.94 &  2,915,802.29 &   0.013729 \\
        & 50\%  &  8,237,387.50 &  53,383,376,977.72 &  2,006,091.50 &   14,073,439,429.50 &  3,384,654.11 &   0.014579 \\
        & 75\%  &  9,405,905.75 &  59,188,646,444.18 &  2,398,085.25 &   15,973,362,072.81 &  4,050,343.31 &   0.015660 \\
        & max   &    15,909,358 &  99,048,039,796.82 &     4,642,419 &   27,685,776,913.57 &  9,097,891.31 &   0.032760 \\
        \midrule
        \multirow{7}{*}{Dow 30$^{\prime}$}   & mean  &  1,555,958.65 &  15,313,345,373.03 &    346,953.30 &    3,573,555,257.78 &    635,769.54 &   0.011792 \\
        & std   &    463,558.93 &   3,891,299,900.31 &    146,677.85 &    1,234,882,079.43 &    655,911.15 &   0.008071 \\
        & min   &       579,206 &   6,664,671,053.15 &        89,564 &    1,035,855,029.71 &    145,205.65 &   0.008879 \\
        & 25\%  &  1,278,813.25 &  12,915,031,172.08 &       262,209 &    2,804,569,367.64 &    417,485.73 &   0.009667 \\
        & 50\%  &     1,429,062 &  14,431,597,662.01 &       309,158 &    3,274,390,601.60 &    514,856.64 &   0.010213 \\
        & 75\%  &  1,715,351.25 &  16,829,521,684.38 &       387,772 &    3,993,470,514.97 &    666,268.27 &   0.011288 \\
        & max   &     3,596,006 &  30,999,914,293.66 &     1,073,029 &    9,428,952,387.10 &  7,817,684.58 &   0.093108 \\
        \bottomrule
    \end{tabular}}
    \label{tab:purse-eyecatchers}
\end{table}

\begin{table}
    \tbl{
    Skew and kurtosis for daily ROC by mutually-exclusive market category, highlighting the remarkably heavy-tailed nature of these distributions.
    }
    {\begin{tabular}{lrr}
        \toprule
        \tabhead{}{Skew}{Kurtosis}
        \midrule
        Dow     &    52.59 &   3122.65\\
        SPexDow &    55.66 &   5644.74 \\
        RexSP   &   300.12 & 110365.89 \\
        \bottomrule
    \end{tabular}}
    \label{tab:bucket-daily-roc-kurtosis-skew}
\end{table}

\begin{table}
    \tbl{
    Summary statistics for realized opportunity cost (ROC) observed in the ETFs under study.
    It is notable that, of all market subsets we study, only this small subset has a ratio of the fraction of differing traded value to fraction of differing trades with value below unity.
    On a per-trade basis, this means that there is on average less potential for ROC.
    }
    {\begin{tabular}{llr}
        \toprule
        1 & Realized Opportunity Cost   &         \$38,458,070.79\\
        2 & SIP Opportunity Cost        &        \$37,970,135.30   \\
        3 & Direct Opportunity Cost     &           \$487,935.49   \\
        4 & Trades                      &          86,725,286      \\
        5 & Diff.\ Trades               &          19,612,214      \\
        6 & Traded Value                & \$3,678,242,397,422.43   \\
        7 & Diff.\ Traded Value         &   \$804,917,872,051.93   \\
        8 & Percent Diff.\ Trades       &                  22.61   \\
        9 & Percent Diff.\ Traded Value &                  21.88   \\
        10 & Ratio of 9 / 8             &                   0.9677 \\
        \bottomrule
    \end{tabular}}
    \label{tab:etfs-eyecatchers}
\end{table}

\begin{table}
    \tbl{
    Aggregated purse statistics for the ETFs under study.
    The data used to construct this table is aggregated by date and instrument, resulting in 2,259 data points that correspond with the 2,268 combinations of 252 trading days in 2016 and 9 ETFs under study.
    }
    {\begin{tabular}{lrrrrrr}
        \toprule
        \tabhead{}{Trades}{Traded Value (\$)}{Diff.\ Trades}{Diff.\ Traded Value (\$)}{ROC (\$)}{ROC/Share}
        \midrule
        mean  &  38,391.01 &  1,628,261,353.44 &      8,681.81 &      356,316,012.42 &  17,024.38 &   0.021169 \\
        std   & 106,302.46 &  4,663,474,508.49 &     23,900.69 &    1,033,570,406.20 &  48,481.79 &   0.043449 \\
        min   &          1 &           72.4600 &             0 &                   0 &          0 &          0 \\
        25\%  &         14 &        262,574.18 &             3 &           48,125.50 &    35.0000 &   0.008350 \\
        50\%  &        683 &     15,165,081.37 &           181 &        3,386,159.33 &   455.2200 &   0.009997 \\
        75\%  &  12,121.50 &    283,540,074.38 &         4,136 &       93,960,790.38 &   6,033.43 &   0.014408 \\
        max   &    974,888 & 40,617,035,891.21 &       251,657 &   11,028,368,359.92 & 499,906.77 &     1.0200 \\
        \bottomrule
    \end{tabular}}
    \label{tab:etf-day-ticker}
\end{table}

\begin{table}
    \tbl{
    Aggregated purse statistics for the ETFs under study.
    The data used to construct this table is aggregated by date, resulting in 252 data points that correspond with the 252 trading days in 2016.
    }
    {\begin{tabular}{lrrrrrr}
        \toprule
        \tabhead{}{Trades}{Traded Value (\$)}{Diff.\ Trades}{Diff.\ Traded Value (\$)}{ROC (\$)}{ROC/Share}
        \midrule
        mean  & 344,147.96 & 14,596,199,989.77 &     77,826.25 &    3,194,118,539.89 & 152,611.39 &   0.189762\\
        std   & 157,107.76 &  6,043,079,696.41 &     45,179.00 &    1,675,731,349.39 &  85,509.19 &   0.118446 \\
        min   &    113,860 &  5,018,912,183.01 &        14,610 &      703,559,994.91 &  30,989.52 &   0.054358 \\
        25\%  & 237,021.25 & 10,471,387,904.01 &     47,237.50 &    2,052,459,478.17 &  94,488.20 &   0.106098 \\
        50\%  &    308,705 & 13,005,695,875.47 &        66,509 &       2,780,132,908 & 131,084.42 &   0.169572 \\
        75\%  & 394,822.25 & 16,641,275,220.96 &        94,108 &    3,799,483,257.76 & 186,174.78 &   0.256871 \\
        max   &  1,177,148 & 44,900,644,748.00 &       339,480 &   12,945,336,256.63 & 616,859.86 &     1.0963 \\
        \bottomrule
    \end{tabular}}
    \label{tab:etf-day}
\end{table}
\clearpage
\pagebreak

\section{Figures}\label{sec:figures}

\begin{figure}
    \centering
    \includegraphics[width=0.8\textwidth]{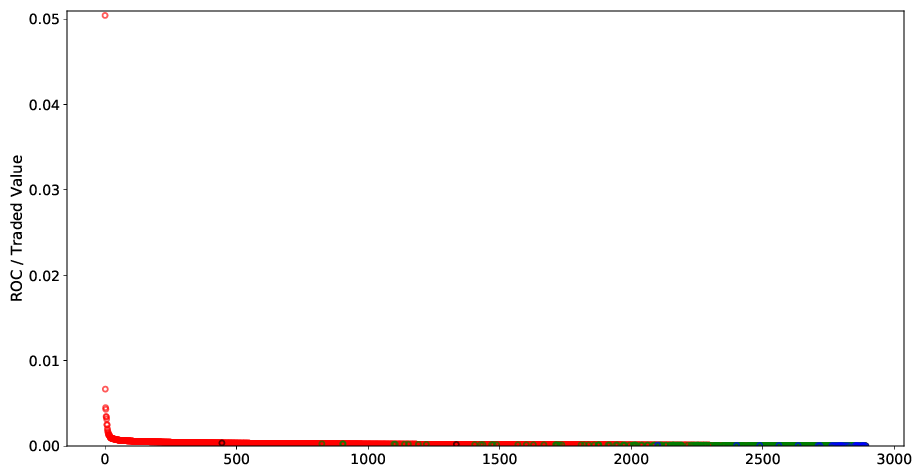}
    \caption{
    Equities are plotted in rank-order of ROC per traded value;
    the 0-th equity has highest ROC per traded value.
    The first over-100 top equities are in the RexSP, which is unsurprising due to their combination of generally lower liquidity and lower share prices.
    Blue markers are associated with constituents of the Dow 30, green markers with constituents of the S\&P 500 (excluding the Dow 30), red markers with constituents of the Russell 3000 (excluding the S\&P 500), and black markers with ETFs.
    }
    \label{fig:roc-traded-value}
\end{figure}

\begin{figure}
    \centering
    \includegraphics[width=0.8\textwidth]{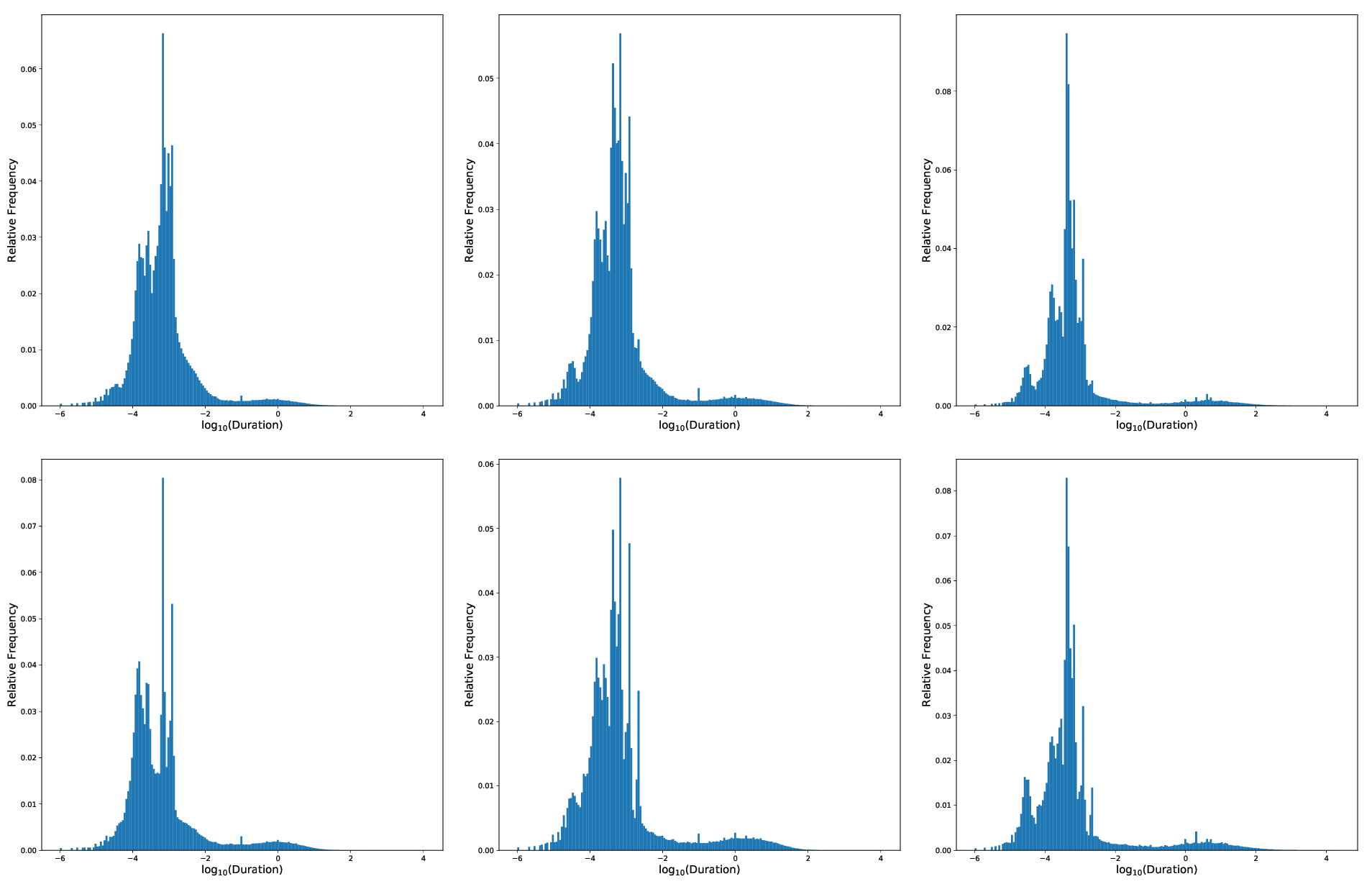}
    \caption{
    Distributions of dislocation segment duration.
    Columns are associated with an index (left to right: Dow 30, S\&P 500 excluding the Dow 30, Russell 3000 excluding the S\&P 500) and rows are associated with conditioning strategies (top to bottom: no conditioning, magnitude greater than 1\cent).
    }
    \label{fig:duration_histograms}
\end{figure}

\begin{figure}
    \centering
    \includegraphics[width=0.8\textwidth]{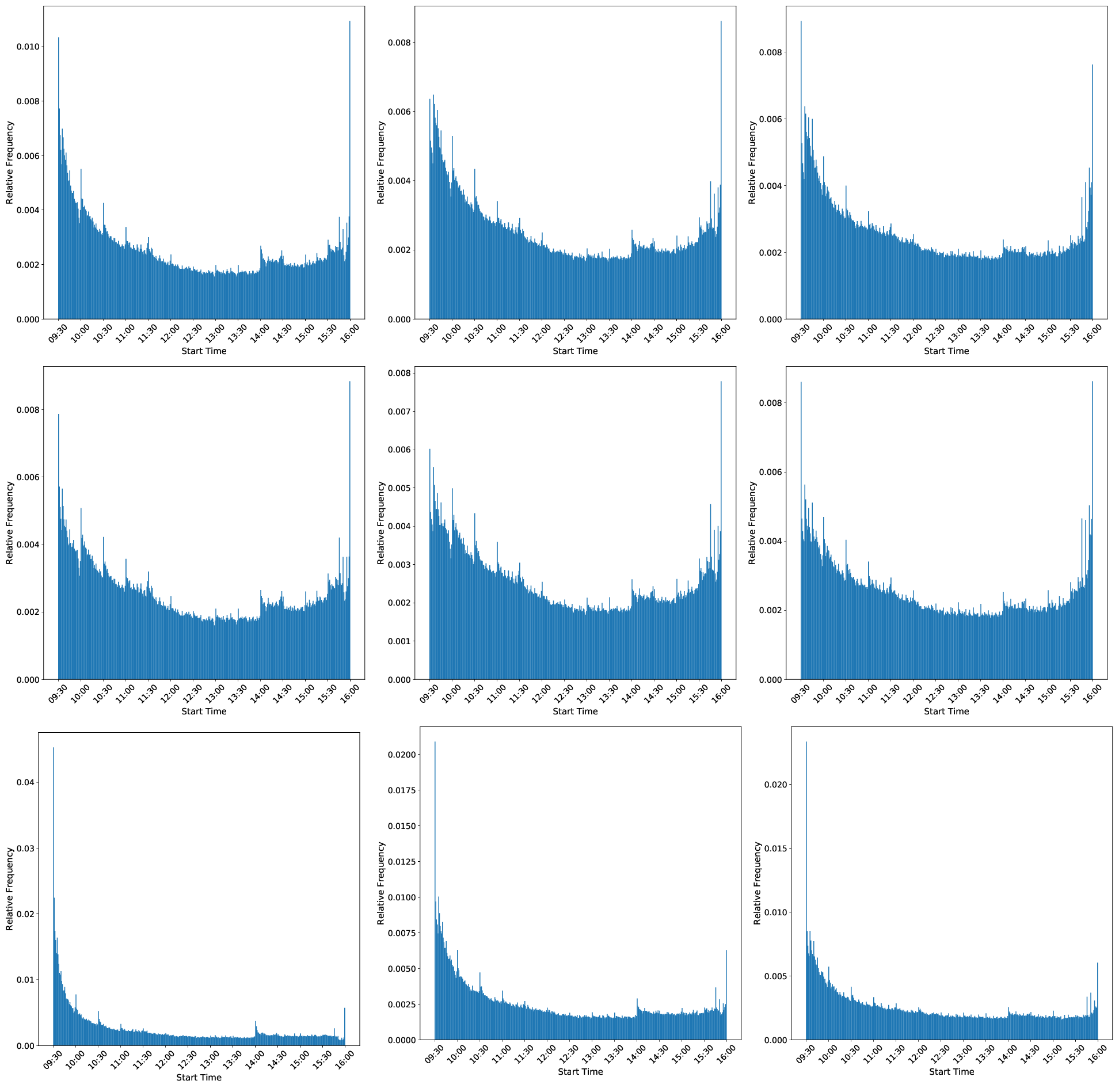}
    \caption{
    Distributions of dislocation segment start time.
    Columns are associated with an index (left to right: Dow 30, S\&P 500 excluding the Dow 30, Russell 3000 excluding the S\&P 500) and rows are associated with conditioning strategies (top to bottom: no conditioning, duration greater than 545$\mu s$, duration greater than 545$\mu s$ and magnitude greater than 1\cent).
    }
    \label{fig:start_histograms}
\end{figure}
\clearpage
\pagebreak

\section{Statistics}\label{sec:statistics}

\begin{table}
    \centering
    \tbl{
    Granger causality results for pairwise combinations of mutually-exclusive market category under study.
    Statistical significance was assessed using four Granger causality tests (parameter $F$-test, sum of squared residuals $F$-test, likelihood-ratio test, $\chi^2$-test).
    Each causal relationship was considered significant if each of the four tests resulted in a $p$-value less than $0.05 / N_{\text{lags}}$.
    The maximum number of lags investigated was $N_{\text{lags}} = 40$.
    }
    {\begin{tabular}{ll}
        \toprule
        \tabhead{\textbf{Ordered pair}}{\textbf{Lags}}
        \midrule
        Dow     $\rightarrow$ RexSP   & 2, 3, 4, 13, 14, 15, 20, 22, \ldots, 37\\
        Dow     $\rightarrow$ SPexDow &                                    ~\\
        SPexDow $\rightarrow$ Dow     & 1, \ldots, 10, 15,\ldots,24, 26, 30, \ldots, 34  \\
        SPexDow $\rightarrow$ RexSP   & 1, 2, 3, 4                           \\
        RexSP   $\rightarrow$ Dow     & 1, 3, 35, 36 \\
        RexSP   $\rightarrow$ SPexDow &                                    ~\\
        \bottomrule
    \end{tabular}}
    \label{tab:granger-lags}
\end{table}

\begin{table}
    \centering
    \tbl{Ordinary least squares regression predicting realized opportunity cost (ROC) using market capitalization, differing trades, and total trades.}{}
    \begin{tabular}{lclc}
        \toprule
        \textbf{Dep. Variable:}       & $\log_{10}$ ROC & \textbf{  R-squared:         } &     0.908   \\
        \textbf{Model:}               &      OLS        & \textbf{  Adj. R-squared:    } &     0.908   \\
        \textbf{Method:}              & Least Squares   & \textbf{  F-statistic:       } &     7179.   \\
        \textbf{No. Observations:}    &       2884      & \textbf{  Prob (F-statistic):} &     0.00    \\
        \textbf{Df Residuals:}        &       2880      & \textbf{  Log-Likelihood:    } &    551.07   \\
        \textbf{Df Model:}            &          3      & \textbf{  AIC:               } &    -1094.   \\
        &                 & \textbf{  BIC:               } &    -1070.   \\
        \textbf{Omnibus:}       & 1630.431 & \textbf{  Durbin-Watson:     } &     2.007  \\
        \textbf{Prob(Omnibus):} &   0.000  & \textbf{  Jarque-Bera (JB):  } & 23812.396  \\
        \textbf{Skew:}          &   2.375  & \textbf{  Prob(JB):          } &      0.00  \\
        \textbf{Kurtosis:}      &  16.252  & \textbf{  Cond. No.          } &      259.  \\
        \bottomrule
    \end{tabular}
    \begin{tabular}{lcccccc}
        & \textbf{coef} & \textbf{std err} & \textbf{z} & \textbf{P$>$$|$z$|$} & \textbf{[0.025} & \textbf{0.975]}  \\
        \midrule
        \textbf{Intercept}            &       1.0052  &        0.091     &    11.050  &         0.000        &        0.827    &        1.183     \\
        \textbf{l\_MarketCap}         &       0.1183  &        0.011     &    10.675  &         0.000        &        0.097    &        0.140     \\
        \textbf{l\_total\_trades}     &      -0.2203  &        0.043     &    -5.127  &         0.000        &       -0.304    &       -0.136     \\
        \textbf{l\_differing\_trades} &       0.9023  &        0.040     &    22.286  &         0.000        &        0.823    &        0.982     \\
        \bottomrule
    \end{tabular}
    \label{tab:roc-market-cap}
\end{table}

\begin{table}
    \centering
    \tbl{
    Ordinary least squares regression predicting realized opportunity cost (ROC) using market capitalization, differing trades, and total trades.
    Quadratic terms are included.
    }{}
    \begin{tabular}{lclc}
        \toprule
        \textbf{Dep. Variable:}                    &    $\log_{10}$ ROC     & \textbf{  R-squared:         } &     0.925   \\
        \textbf{Model:}                            &       OLS        & \textbf{  Adj. R-squared:    } &     0.925   \\
        \textbf{Method:}                           &  Least Squares   & \textbf{  F-statistic:       } &     5970.   \\
        \textbf{No. Observations:}                 &        2884      & \textbf{  Prob (F-statistic):} &     0.00    \\
        \textbf{Df Residuals:}                     &        2877      & \textbf{  Log-Likelihood:    } &    846.73   \\
        \textbf{Df Model:}                         &           6      & \textbf{  AIC:               } &    -1679.   \\
        &                  & \textbf{  BIC:               } &    -1638.   \\
        \textbf{Omnibus:}       & 1952.210 & \textbf{  Durbin-Watson:     } &     1.988  \\
        \textbf{Prob(Omnibus):} &   0.000  & \textbf{  Jarque-Bera (JB):  } & 50808.169  \\
        \textbf{Skew:}          &   2.831  & \textbf{  Prob(JB):          } &      0.00  \\
        \textbf{Kurtosis:}      &  22.768  & \textbf{  Cond. No.          } &  1.70e+04  \\
        \bottomrule
    \end{tabular}

    \begin{tabular}{lcccccc}
        & \textbf{coef} & \textbf{std err} & \textbf{z} & \textbf{P$>$$|$z$|$} & \textbf{[0.025} & \textbf{0.975]}  \\
        \midrule
        \textbf{Intercept}                         &       7.8666  &        0.802     &     9.811  &         0.000        &        6.295    &        9.438     \\
        \textbf{l\_MarketCap}                      &      -0.0738  &        0.149     &    -0.496  &         0.620        &       -0.365    &        0.218     \\
        \textbf{l\_total\_trades}                  &      -4.1661  &        0.432     &    -9.638  &         0.000        &       -5.013    &       -3.319     \\
        \textbf{l\_differing\_trades}              &       3.0804  &        0.338     &     9.103  &         0.000        &        2.417    &        3.744     \\
        \textbf{l\_MarketCap ** 2}         &       0.0067  &        0.008     &     0.837  &         0.402        &       -0.009    &        0.022     \\
        \textbf{l\_total\_trades ** 2}     &       0.3385  &        0.038     &     8.936  &         0.000        &        0.264    &        0.413     \\
        \textbf{l\_differing\_trades ** 2} &      -0.2042  &        0.034     &    -6.002  &         0.000        &       -0.271    &       -0.138     \\
        \bottomrule
    \end{tabular}
    \label{tab:roc-market-cap-quad}
\end{table}

\begin{table}
    \centering
    \tbl{Ordinary least squares regression predicting realized opportunity cost (ROC) using only market capitalization.}{}
    \begin{tabular}{lclc}
        \toprule
        \textbf{Dep. Variable:}    &    $\log_{10}$ ROC    & \textbf{  R-squared:         } &     0.600   \\
        \textbf{Model:}            &       OLS        & \textbf{  Adj. R-squared:    }      &     0.600   \\
        \textbf{Method:}           &  Least Squares   & \textbf{  F-statistic:       }      &     4280.   \\
        \textbf{No. Observations:} &        2884      & \textbf{  Prob (F-statistic):}      &     0.00    \\
        \textbf{Df Residuals:}     &        2882      & \textbf{  Log-Likelihood:    }      &   -1574.9   \\
        \textbf{Df Model:}         &           1      & \textbf{  AIC:               }      &     3154.   \\
        &                  & \textbf{  BIC:               }      &     3166.   \\
        \textbf{Omnibus:}          & 52.492           & \textbf{  Durbin-Watson:     }      &    1.933  \\
        \textbf{Prob(Omnibus):}    &  0.000           & \textbf{  Jarque-Bera (JB):  }      &   76.592  \\
        \textbf{Skew:}             &  0.199           & \textbf{  Prob(JB):          }      & 2.34e-17  \\
        \textbf{Kurtosis:}         &  3.692           & \textbf{  Cond. No.          }      &     126.  \\
        \bottomrule
    \end{tabular}
    \begin{tabular}{lcccccc}
        & \textbf{coef} & \textbf{std err} & \textbf{z} & \textbf{P$>$$|$z$|$} & \textbf{[0.025} & \textbf{0.975]}  \\
        \midrule
        \textbf{Intercept}    &      -1.4415  &        0.108     &   -13.398  &         0.000        &       -1.652    &       -1.231     \\
        \textbf{l\_MarketCap} &       0.7368  &        0.011     &    65.422  &         0.000        &        0.715    &        0.759     \\
        \bottomrule
    \end{tabular}
    \label{tab:roc-market-cap-marginal}
\end{table}

\begin{table}
    \centering
    \tbl{
    Ordinary least squares regression predicting realized opportunity cost (ROC) using only market capitalization.
    Quadratic terms are included.
    }{}
    \begin{tabular}{lclc}
        \toprule
        \textbf{Dep. Variable:}            &    $\log_{10}$ ROC    & \textbf{  R-squared:         } &     0.603   \\
        \textbf{Model:}                    &       OLS        & \textbf{  Adj. R-squared:    } &     0.603   \\
        \textbf{Method:}                   &  Least Squares   & \textbf{  F-statistic:       } &     2904.   \\
        \textbf{No. Observations:}         &        2884      & \textbf{  Prob (F-statistic):} &     0.00    \\
        \textbf{Df Residuals:}             &        2881      & \textbf{  Log-Likelihood:    } &   -1564.7   \\
        \textbf{Df Model:}                 &           2      & \textbf{  AIC:               } &     3135.   \\
        &                  & \textbf{  BIC:               } &     3153.   \\
        \textbf{Omnibus:}       & 67.584 & \textbf{  Durbin-Watson:     } &    1.935  \\
        \textbf{Prob(Omnibus):} &  0.000 & \textbf{  Jarque-Bera (JB):  } &  100.782  \\
        \textbf{Skew:}          &  0.242 & \textbf{  Prob(JB):          } & 1.30e-22  \\
        \textbf{Kurtosis:}      &  3.777 & \textbf{  Cond. No.          } & 1.24e+04  \\
        \bottomrule
    \end{tabular}
    \begin{tabular}{lcccccc}
        & \textbf{coef} & \textbf{std err} & \textbf{z} & \textbf{P$>$$|$z$|$} & \textbf{[0.025} & \textbf{0.975]}  \\
        \midrule
        \textbf{Intercept}                 &      -6.2441  &        1.286     &    -4.857  &         0.000        &       -8.764    &       -3.724     \\
        \textbf{l\_MarketCap}              &       1.7575  &        0.266     &     6.598  &         0.000        &        1.235    &        2.280     \\
        \textbf{l\_MarketCap ** 2} &      -0.0539  &        0.014     &    -3.927  &         0.000        &       -0.081    &       -0.027     \\
        \bottomrule
    \end{tabular}
    \label{tab:roc-market-cap-marginal-quad}
\end{table}

\end{document}